\pdfoutput=1
\documentclass[aps,prl,reprint,amsmath,amssymb,longbibliography,preprintnumbers]{revtex4-2}
\usepackage[T1]{fontenc}
\usepackage{graphicx}
\usepackage{bm}
\usepackage{mathrsfs}
\usepackage[hidelinks]{hyperref}
\hypersetup{pdftitle={The Schwarzschild-Kerr crossover of high overtone quasinormal modes},
 pdfauthor={Yasuyuki Hatsuda},pdfsubject={Main Letter and Supplemental Material}}
\graphicspath{{figures/}}
\newcommand{\F}{\mathcal F}
\newcommand{\dd}{\mathrm d}
\newcommand{\eps}{\varepsilon}
\newcommand{\ii}{\mathrm i}
\newcommand{\eqrefp}[1]{Eq.~\eqref{#1}}
\begin{document}
\preprint{RUP-26-34}
\title{The Schwarzschild--Kerr crossover of high overtone quasinormal modes}
\author{Yasuyuki Hatsuda}
\affiliation{Department of Physics, Rikkyo University, Toshima, Tokyo 171-8501, Japan}
\begin{abstract}
A slowly rotating black hole is close to a nonrotating one, but its highly damped quasinormal spectrum is not.
The Schwarzschild approximation holds only up to overtone numbers of order $\chi^{-4/3}$, where $\chi$ is the dimensionless spin.
We find that the frequency shift at this scale is a universal function, derived from the wave boundary conditions with an error estimate.
It is given by the spectrum of a sextic oscillator, via the correspondence between ordinary differential equations and integrable models.
It recovers the Schwarzschild $\log 3$ and organizes the singular Kerr coefficients.
\end{abstract}
\maketitle

\textit{Introduction.}
Quasinormal modes (QNMs) are the damped oscillations of a perturbed black hole~\cite{KokkotasSchmidt,BertiCardosoStarinets,KonoplyaZhidenko}.
Their frequencies are labeled by the overtone number, and the damping grows with it.
For a Schwarzschild black hole of mass $M$, the highly damped frequencies of scalar and gravitational perturbations approach~\cite{Nollert,Motl,MotlNeitzke,MusiriSiopsis}
\begin{equation}
2M\omega_N\simeq\frac{\log3}{4\pi}-\frac{\ii}{2}\left(N+\frac12\right),
\label{eq:schw}
\end{equation}
where $N$ is a large integer.
These frequencies are determined by the wave equation near the singularity~\cite{MotlNeitzke,FestucciaLiu}, and the highly damped modes have recently been used as probes of black hole singularities~\cite{HartnollZhiboedov,GrozdanovMovrinValach,CovielloEtAl,JiaRangamani}.
For a Kerr black hole, the highly damped spectrum looks different~\cite{BertiKerr,BertiYoshida,KeshetHod,KeshetNeitzke}.
Its real part is proportional to the azimuthal number $m$, however slow the rotation is, while the real part of Eq.~\eqref{eq:schw} is the same for all $m$.
The two limits, zero rotation and large damping, do not commute~\cite{BertiKerr}.

At a fixed overtone number, a slow rotation is a small perturbation~\cite{Chen}, but the correction grows with the damping~\cite{BertiYoshida,KeshetHod,KeshetNeitzke,DaghighGreenMulligan}.
How slow must the rotation be for Eq.~\eqref{eq:schw} to remain valid?
Earlier works studied the regime $|a\omega|\lesssim1$, where $a$ is the rotation parameter~\cite{MusiriSiopsisKerr}.
Daghigh, Green, and Mulligan studied the transition by complex WKB methods, and found a change around $|a\omega|\sim1$~\cite{DaghighGreenMulligan}.
A similar transition was found in the small charge limit of Reissner--Nordstr\"om black holes~\cite{DaghighRN}.
It is also known that some coefficients of the Kerr expansion at fixed rotation are singular when the rotation vanishes~\cite{KeshetHod,KeshetNeitzke,Paper1}.
These coefficients restrict the expansion at fixed rotation to $N\gg(a/M)^{-4/3}$~\cite{Paper1}.
The crossover between the two highly damped spectra has remained an open problem.

In this Letter, we show that the crossover takes place at $N\sim(a/M)^{-4/3}$, and we determine the function that replaces $\log3$.
We use the dimensionless frequency $\Omega$ and the dimensionless spin $\chi$,
\begin{equation}
\Omega=2M\omega,\qquad \chi=\frac aM,\qquad
w_N=\frac12\left(N+\frac12\right),
\label{eq:scales}
\end{equation}
so that Eq.~\eqref{eq:schw} reads $\Omega_N\simeq\log3/(4\pi)-\ii w_N$.
We take $N\to\infty$ and $\chi\to0$ with $\xi=\chi w_N^{3/4}$ fixed.
Our main result is
\begin{equation}
\Omega_N=-\ii w_N+\F_{0,s}(\xi)+O(\eps),\qquad \eps=w_N^{-1/4}.
\label{eq:expansion}
\end{equation}
The Schwarzschild formula thus breaks down when $N$ is of order $\chi^{-4/3}$.
For example, $\xi=1$ corresponds to $N\simeq43$ for $\chi=0.1$, and to $N\simeq930$ for $\chi=0.01$.
The effect of a slow rotation is therefore controlled by $\chi N^{3/4}$ rather than by $\chi$ alone.
We call $\F_{0,s}$ the crossover function.
The integer $N$ agrees with the overtone number for the modes whose corrections at finite $N$ are examined below.

\begin{figure*}[t]
\centering
\includegraphics[width=0.485\textwidth]{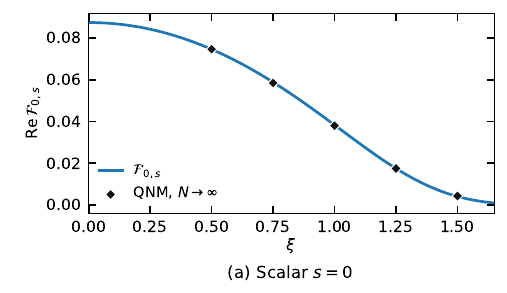}\hfill
\includegraphics[width=0.485\textwidth]{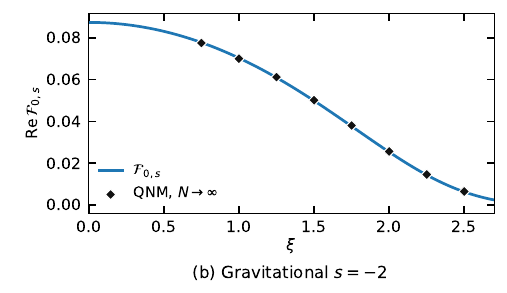}\\[2pt]
\includegraphics[width=0.485\textwidth]{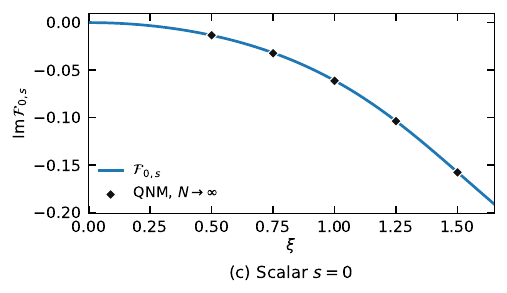}\hfill
\includegraphics[width=0.485\textwidth]{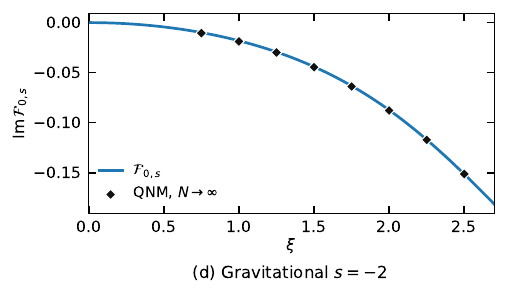}
\caption{The crossover function $\F_{0,s}(\xi)$ (curves) and the limit of $\Omega_N+\ii w_N$ as $N\to\infty$ at fixed $\xi$ (points).
Here $\Omega_N$ is the QNM frequency of the Teukolsky equation at $\chi=\xi w_N^{-3/4}$.
The points are obtained by extrapolation from $10^3\le N\le6.4\times10^5$.
The scalar modes have $(s,\ell,m)=(0,0,0)$, and the gravitational modes have $(-2,2,-2)$.
The curves are parameter-free predictions, computed from the spectral determinant of the sextic oscillator through Eqs.~\eqref{eq:main}, \eqref{eq:Q}, and \eqref{eq:Tproduct}.
The error bars show half the range of five extrapolations, described in the Supplemental Material.
Most of them are smaller than the points.}
\label{fig:crossover}
\end{figure*}

The crossover function depends on the perturbation only through its spin weight $s$.
It is the same for all angular numbers $\ell$ and $m$ on the normal branches introduced below, and in this sense it is universal.
We consider scalar ($s=0$) and gravitational ($s=-2$) perturbations.
As shown below, $\F_{0,s}$ is determined by the spectrum of a sextic oscillator, through the correspondence between ordinary differential equations and integrable models~\cite{DoreyTateoEarly,BLZspec,DoreyReview}.
It equals $\log3/(4\pi)$ at $\xi=0$, while its real part is exponentially small at large $\xi$.
The $\log3$ therefore disappears beyond the crossover (see Fig.~\ref{fig:crossover}).
The imaginary part reproduces the singular coefficients of the Kerr expansion at fixed rotation, whose values are given by the conserved charges of an integrable model.

The crossover scale $N\sim\chi^{-4/3}$ is larger than the scale $N\sim\chi^{-1}$, where $|a\omega|\sim1$ and the angular eigenfunctions change their character~\cite{BertiCardosoCasals,VickersCook}.
At the crossover, $|a\omega|$ is of order $\chi^{-1/3}$, and the angular transition has already taken place.
The crossover is therefore controlled by the radial equation.

We derive Eq.~\eqref{eq:expansion} from the QNM boundary conditions for frequencies with positive real part.
The error estimate is given in the Supplemental Material~\cite{Supplemental}.
We also compare the result with numerical QNM frequencies.
The corrections are of order $N^{-1/4}$, and we identify their main source.

\textit{Scaling limit.}
We use the time dependence $e^{-\ii\omega t+\ii m\phi}$.
The angular equation gives a separation constant $A(c)$, which depends on the rotation through $c=a\omega=\chi\Omega/2$.
We follow $A(c)$ continuously from its value at $c=0$, which is $A(0)=\ell(\ell+1)-s(s+1)$.
For highly damped modes, $c$ is large and almost imaginary.
We consider the branches on which $A(c)$ grows linearly in $c$ in this regime~\cite{VickersCook,BertiCardosoCasals}.
Following Ref.~\cite{VickersCook}, we call them normal branches.
The precise conditions are stated in the Supplemental Material.

For the radial part, we use an alternative~\cite{HatsudaAlternative} to the Teukolsky equation~\cite{Teukolsky},
\begin{equation}
\begin{gathered}
\left[f\frac{\dd}{\dd z}f\frac{\dd}{\dd z}+\Omega^2-V(z)\right]\phi=0,
\qquad f=1-\frac1z,\\
V=f\left[4c^2+\frac{4c(m-c)}z+\frac B{z^2}+\frac{1-s^2}{z^3}\right],
\end{gathered}
\label{eq:radial}
\end{equation}
where $B=A(c)+s(s+1)-2mc+c^2$.
Here $z$ is an auxiliary radial variable.
The event horizon is at $z=1$, and $z=0$ is an inner singular point.
In the regime of interest, Eq.~\eqref{eq:radial} has the same QNM frequencies as the Teukolsky equation, as we show in the Supplemental Material.
An advantage of Eq.~\eqref{eq:radial} is that the scaling limit is very simple.

As in the Schwarzschild case~\cite{MotlNeitzke}, the highly damped QNMs are controlled by the neighborhood of a singular point, which is $z=0$ here.
For $y=\sqrt f\,\phi$, Eq.~\eqref{eq:radial} reads $y''+[\Omega^2z^2+B/z+(1/4-s^2)/z^2]y=0$ near $z=0$, up to smaller terms.
These three terms compete there.
They are the frequency term $\Omega^2z^2$, the term $(1/4-s^2)/z^2$, which is already present for Schwarzschild black holes, and the term $B/z$, which carries the rotation.
The first two terms balance for $z$ of order $|\Omega|^{-1/2}$, and the rotation term is equally important if $B$ is of order $|\Omega|^{1/2}$.
Since $B\simeq c^2=\chi^2\Omega^2/4$ on the normal branches, this happens for $\chi|\Omega|^{3/4}$ of order one.
This is the origin of the variable $\xi$.
In terms of $v=-\ii\sqrt\Omega\,z$, where $\sqrt\Omega\simeq e^{-\ii\pi/4}\sqrt{w_N}$, it becomes
\begin{equation}
\frac{\dd^2y}{\dd v^2}+\left[v^2+\frac{\kappa_0}{v}
+\frac{1/4-s^2}{v^2}\right]y=0,
\qquad \kappa_0=\frac{e^{-\ii\pi/4}}4\xi^2
\label{eq:inner}
\end{equation}
at leading order.
For $\xi=0$, it reduces to a Bessel equation, as in the Schwarzschild problem~\cite{MotlNeitzke}.
The rotation adds a Coulomb-like term, whose strength is $\xi^2$.
Equation~\eqref{eq:inner} depends on the perturbation only through the spin weight $s$.
The angular numbers $\ell$ and $m$ first appear at order $\eps$, which explains the universality of $\F_{0,s}$.

In practice, the corrections of order $\eps$ are sizable.
At this order, they come from the terms of $B$ linear in $c$, which are the linear term of $A(c)$ and $-2mc$.
We include them by replacing $\kappa_0$ in Eq.~\eqref{eq:inner} with
\begin{equation}
\kappa_{\rm eff}=\frac{\ii B}{\sqrt\Omega}=\kappa_0+O(\eps),
\label{eq:kappa}
\end{equation}
where $B$ is evaluated with the exact $A(c)$.
We use $\kappa_{\rm eff}$ in the numerical tests below.

\textit{Quantization condition.}
For Schwarzschild black holes, the asymptotic QNM condition follows from a monodromy argument~\cite{MotlNeitzke,AnderssonHowls}.
One continues the outgoing wave around the horizon in the complex radial plane, and compares it with the solution near the singular point.
The result is $e^{4\pi\Omega}+3=0$, which gives Eq.~\eqref{eq:schw}.

We apply the same idea to Eq.~\eqref{eq:inner}, with $\kappa_0$ replaced by a complex parameter $\kappa$.
Its solutions behave as $v^{-1/2}e^{\pm\ii v^2/2}$ at large $|v|$.
We continue the wave with the upper sign clockwise around $v=0$ by the angle $3\pi/2$.
On the final ray, we denote the coefficient of the same wave by $-T_s(\kappa)$.
At $\kappa=0$, the Bessel functions give $T_s(0)=1+2\cos(\pi s)$, which is $3$ for $s=0,-2$.
We find that the QNM frequencies satisfy
\begin{equation}
e^{4\pi\Omega}+T_s(\kappa_{\rm eff})
=O(\eps^2).
\label{eq:physicalcondition}
\end{equation}
The number $3$ of the Schwarzschild condition is replaced by a function of the rotation.
Using $\kappa_{\rm eff}=\kappa_0+O(\eps)$ and solving for $\Omega$, we obtain Eq.~\eqref{eq:expansion} with
\begin{equation}
\F_{0,s}(\xi)=\frac1{4\pi}\log T_s(\kappa_0).
\label{eq:main}
\end{equation}
The logarithm is defined by continuity from $\log3$ at $\xi=0$.

Equation~\eqref{eq:physicalcondition} holds when $\operatorname{Re}\Omega$ and $\xi$ lie in fixed ranges of positive values.
We also require that the angular eigenvalue is analytic near the frequencies of interest and stays on a normal branch.
The derivation has two steps.
We first show that Eq.~\eqref{eq:radial} and the Teukolsky equation have the same QNM frequencies under these conditions~\cite{Whiting,CasalsTeixeira,KazakovSlavyanov}.
We then estimate the error of the matching between the solution of Eq.~\eqref{eq:inner} and the outgoing waves.
Conversely, each simple root of $e^{4\pi\Omega}+T_s(\kappa_{\rm eff})=0$ in this region has exactly one QNM frequency within a distance of order $\eps^2$.
The details are given in the Supplemental Material.

\begin{figure*}[t]
\centering
\includegraphics[width=0.485\textwidth]{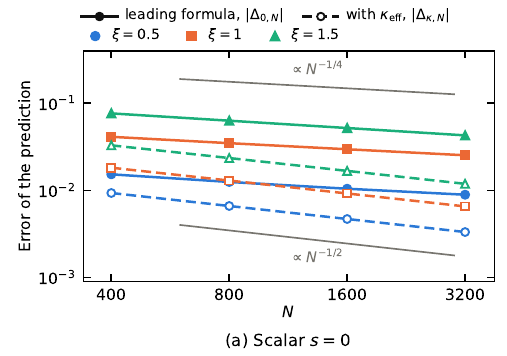}\hfill
\includegraphics[width=0.485\textwidth]{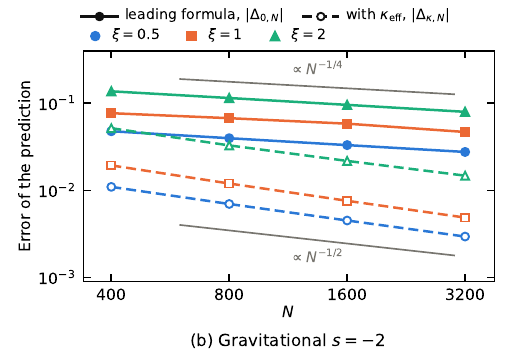}
\caption{Distance between the QNM frequencies $\Omega_N$ of the Teukolsky equation and two predictions at fixed $\xi$.
Solid lines show $|\Delta_{0,N}|$ for the leading formula $-\ii w_N+\F_{0,s}(\xi)$.
Dashed lines show $|\Delta_{\kappa,N}|$ for the root of $e^{4\pi\Omega}+T_s(\kappa_{\rm eff})=0$.
The lines connect the data points.
The gray lines show the slopes $N^{-1/4}$ and $N^{-1/2}$.
The numerical uncertainty of the frequencies is much smaller than the plotted errors.}
\label{fig:errors}
\end{figure*}

\textit{Sextic oscillator.}
The function $T_s(\kappa)$ can be computed from a simple problem in quantum mechanics.
With $v=e^{\ii\pi/4}x^2/\sqrt2$, $y=x^{1/2}\varphi$, and $X=e^{\ii\pi/4}\kappa$, Eq.~\eqref{eq:inner} becomes
\begin{equation}
\left[-\frac{\dd^2}{\dd x^2}+x^6+\frac{4s^2-1/4}{x^2}\right]\varphi
=E\varphi,
\quad E=2\sqrt2X.
\label{eq:sextic}
\end{equation}
This is the Schr\"odinger equation of a sextic oscillator, where $x$ is an auxiliary variable~\cite{DoreyTateoEarly,DoreyTateo,DoreyDunningTateo}.
We consider Eq.~\eqref{eq:sextic} on the half line $x>0$, and require that $\varphi$ behaves as $x^{2|s|+1/2}$ at $x=0$ and decays at infinity.
The eigenvalues are then discrete and positive.
We write them as $E_k=2\sqrt2\lambda_k$ with $k=0,1,2,\ldots$, and define
\begin{equation}
Q_s(X)=\prod_{k\ge0}\left(1-\frac X{\lambda_k}\right).
\label{eq:Q}
\end{equation}
The correspondence between ordinary differential equations and integrable models~\cite{DoreyTateoEarly,BLZspec,DoreyTateo,DoreyDunningTateo,DoreyReview} relates $T_s$ to $Q_s$.
For $s=0,-2$, we find
\begin{equation}
\begin{split}
T_s(e^{-\ii\pi/4}X)+1
={}&\left[1+\frac{Q_s(X)}{Q_s(-X)}\right]\\
&\times\left[1+\frac{Q_s(\ii X)}{Q_s(-\ii X)}\right].
\end{split}
\label{eq:Tproduct}
\end{equation}
In Eq.~\eqref{eq:main}, we need $X=\xi^2/4$, which is real and positive.
The crossover function of the complex QNM frequencies is thus fixed by the real spectrum of the oscillator.
QNMs of D3-branes and of their intersections were studied with the functional relations and integral equations of this correspondence~\cite{FioravantiGregori}.
There the QNM frequencies are the zeros of a $Q$ function.
Here $Q_s$ vanishes at the eigenvalues of the sextic oscillator, and the QNM frequencies follow from $T_s$ through Eq.~\eqref{eq:physicalcondition}.
Related determinant formulas were recently derived for the Stokes multipliers near the singularity of anti-de Sitter black holes, in a limit where the momentum is scaled with the frequency~\cite{JiaRangamani}.
Here two such multipliers combine into $T_s$, and the sextic oscillator determines the crossover of the Kerr QNMs.
The inner limit of the Teukolsky equation gives the same $T_s$ through a related sextic oscillator.
In the scaling limit, the parameters of the gauge theory that describes the Kerr QNMs also approach an Argyres--Douglas point~\cite{AGH_SW,ItoOkubo_AD,ItoKoizumiOkubo_AD}.
At nonzero $\xi$, $\kappa_0$ remains as a deformation of this point.
We explain these two relations in the Supplemental Material.

Let us look at the two limits.
At $\xi=0$, Eq.~\eqref{eq:Tproduct} gives $T_s(0)=3$.
For small $\xi$, we find
\begin{equation}
\F_{0,s}(\xi)=\frac{\log3}{4\pi}
-\frac{(1+\ii)Z_{1,s}}{12\pi}\xi^2+O(\xi^4),
\label{eq:small}
\end{equation}
where $Z_{1,s}=\sum_k\lambda_k^{-1}$ is known in closed form~\cite{Watkins}, given in Eq.~\eqref{supp:Z1} of the Supplemental Material.
Recall that $\xi^2=\chi^2w_N^{3/2}$.
The term of order $\xi^2$ therefore grows as $N^{3/2}$ at fixed $\chi$, and it is of order one when $N$ is of order $\chi^{-4/3}$.

At large $\xi$, the first factor in Eq.~\eqref{eq:Tproduct} tends to one exponentially fast, while the ratio in the second factor has modulus one.
Therefore $|T_s|\to1$, and the real part of $\F_{0,s}$ is exponentially small.
Where it is nonzero, its sign is that of $Q_s(X)$.
The real part first changes sign at $\xi=2\sqrt{\lambda_0}$, where $T_s=-1$.
This value is $1.92$ for $s=0$ and $5.57$ for $s=-2$.
For $s=-2$, the real part also touches zero at $\xi=2\cdot6^{1/4}\simeq3.13$, as shown in the Supplemental Material.
Since Eq.~\eqref{eq:physicalcondition} requires $\operatorname{Re}\Omega>0$, our derivation covers only the values of $\xi$ with $\operatorname{Re}\F_{0,s}>0$.
Up to its exponentially small real part, $\F_{0,s}$ has the asymptotic expansion
\begin{equation}
\F_{0,s}(\xi)\sim-\frac{3\ii B_0}{32\pi}\xi^{4/3}
+\frac{\ii|s|}{4}
+\frac{\ii(1/6-s^2)B_1}{4\pi}\xi^{-4/3}+\cdots,
\label{eq:firstlarge}
\end{equation}
where $B_0$ and $B_1$ are positive constants given in the Supplemental Material.
Figure~\ref{supp:figasym} of the Supplemental Material compares Eqs.~\eqref{eq:small} and \eqref{eq:firstlarge} with $\F_{0,s}$.
Each term has a simple meaning at fixed $\chi$.
The first term is proportional to $\xi^{4/3}=\chi^{4/3}w_N$.
It is linear in $N$, and it changes the spacing of the damping rates.
The last term is proportional to $\xi^{-4/3}=\chi^{-4/3}w_N^{-1}$.
It is a correction of order $1/N$, whose coefficient diverges for $\chi\to0$.
The higher terms are proportional to $\chi^{-4j/3}w_N^{-j}$ with odd $j$.
The large $\xi$ expansion thus reproduces the singular coefficients of the Kerr expansion at fixed rotation~\cite{KeshetHod,KeshetNeitzke,Paper1}.
Their values are given by the conserved charges of the integrable model of Bazhanov, Lukyanov, and Zamolodchikov~\cite{BLZI,BLZII}.
In particular, the term with $j=3$ vanishes, so that the coefficient at order $N^{-3}$ has no term proportional to $\chi^{-4}$.
Its leading term is proportional to $\chi^{-10/3}$~\cite{Paper1}.
This is a comparison of coefficients.
We do not have an error estimate that is uniform at large $\xi$.

\textit{Numerical tests.}
We compare Eq.~\eqref{eq:expansion} with QNM frequencies of the Teukolsky equation, computed by Leaver's continued fraction method~\cite{Leaver,Nollert}.
We consider the scalar modes with $(s,\ell,m)=(0,0,0)$ and the gravitational modes with $(-2,2,-2)$.
Equation~\eqref{eq:expansion} is a parameter-free prediction for these frequencies.

We first test the limit $N\to\infty$ at fixed $\xi$.
For each value of $\xi$ in Fig.~\ref{fig:crossover}, we compute $\Omega_N+\ii w_N$ at 20 or more values of $N$, with $\chi=\xi w_N^{-3/4}$.
These values of $N$ lie between $10^3$ and $6.4\times10^{5}$.
We then extrapolate the data to $N\to\infty$ by polynomial fits in $\eps=w_N^{-1/4}$.
The fits do not use Eq.~\eqref{eq:expansion}.
Figure~\ref{fig:crossover} shows the extrapolated values.
They agree with $\F_{0,s}(\xi)$ within $6\times10^{-4}$ for the scalar modes and within $8\times10^{-4}$ for the gravitational modes.
These differences are comparable to the uncertainty of the extrapolation.

We next look at the corrections at finite $N$.
We take three values of $\xi$ for the scalar and for the gravitational modes, with $N=400,800,1600,3200$.
The rotation $\chi$ ranges from $0.002$ to $0.04$.
Even at $N=3200$, we have $\eps\simeq0.158$, and the corrections of order $\eps$ are sizable.
They are larger for the gravitational modes.

To identify them, we also solve $e^{4\pi\Omega}+T_s(\kappa_{\rm eff})=0$ with the exact angular eigenvalue.
We denote its solution by $\Omega_N^{(\kappa)}$, which we computed independently of the QNM data.
The two errors are
\begin{equation}
\begin{aligned}
\Delta_{0,N}&=\Omega_N+\ii w_N-\F_{0,s}(\xi),\\
\Delta_{\kappa,N}&=\Omega_N-\Omega_N^{(\kappa)}.
\end{aligned}
\label{eq:errors}
\end{equation}
Equations~\eqref{eq:expansion} and \eqref{eq:physicalcondition} bound them by $O(\eps)$ and $O(\eps^2)$.
For concreteness, let us consider the gravitational mode with $\xi=2$ and $N=3200$.
The real part of $\Omega_N+\ii w_N$ is $0.1021$.
The leading formula gives $0.0253$, while $\Omega_N^{(\kappa)}$ gives $0.1064$.
The modulus of the error decreases from $0.080$ to $0.015$.

Figure~\ref{fig:errors} shows the two errors for these 24 frequencies.
They decrease approximately as $N^{-1/4}$ and $N^{-1/2}$, respectively.
These slopes are consistent with the two bounds.
In convergence tests, these frequencies change by less than $7\times10^{-16}$.
In all these cases, $N$ agrees with the overtone number.
See the Supplemental Material for details.

\textit{Concluding remarks.}
In this Letter, we studied the highly damped QNMs of slowly rotating black holes.
We found that the Schwarzschild formula breaks down at $N\sim\chi^{-4/3}$, and that the constant $\log3$ is replaced by a function of $\xi=\chi w_N^{3/4}$ at this scale.
This function is determined by the spectrum of a sextic oscillator.
It recovers $\log3$ at $\xi=0$, and it reproduces the singular coefficients of the Kerr expansion at fixed rotation at large $\xi$.

There are several open problems.
First, the corrections of order $\eps$ are sizable for the overtone numbers studied here.
The first one follows from $\kappa_{\rm eff}$ and is given in the Supplemental Material.
It would be nice to compute the higher corrections systematically.
They depend on $\ell$ and $m$.
Secondly, our tests concern $s=0,-2$ and the normal branches.
It would be interesting to study electromagnetic perturbations, for which $T_s(0)=-1$, and the anomalous branches, on which $A(c)$ grows as $c^2$~\cite{VickersCook}.
Thirdly, $\operatorname{Re}\F_{0,s}$ vanishes at a finite value of $\xi$, where the frequencies approach the negative imaginary axis.
It is desirable to extend the error estimate to this region.
Finally, the sextic oscillator is related to a conformal field theory and to an Argyres--Douglas theory.
At the moment, we do not understand the physical meaning of these relations for black holes.

\textit{Data availability.}
The QNM frequencies used for Fig.~\ref{fig:crossover} and a script that computes the crossover function $\F_{0,s}$ are available as ancillary files.
The other numerical data are available from the author upon reasonable request.

\begin{acknowledgments}
\textit{Acknowledgments.}
The author thanks Alba Grassi for sharing the manuscript of Ref.~\cite{GiorgiGrassi} before submission.
The author used ChatGPT and Codex (GPT-6 Sol, GPT-6 Pro, and GPT-6.2 Astra, OpenAI) and Claude and Claude Code (Claude Opus 5.5 and Claude Fable 5.2, Anthropic) for language editing, for assistance with analytic and numerical calculations, and for code generation.
The author directed these uses and checked all the outputs.
The analytic results were checked by independent derivations and symbolic computations, and the numerical results by independent methods and convergence tests, as described in the Supplemental Material.
This work was partially supported by JSPS KAKENHI Grant Numbers JP23K25790 and JP26K07110.
\end{acknowledgments}

\clearpage
\onecolumngrid
\setcounter{secnumdepth}{2}
\setcounter{section}{0}
\setcounter{equation}{0}
\setcounter{table}{0}
\setcounter{figure}{0}
\makeatletter
\@removefromreset{equation}{section}
\makeatother
\renewcommand{\thesection}{\Alph{section}}
\renewcommand{\thesubsection}{\arabic{subsection}}
\renewcommand{\theequation}{S\arabic{equation}}
\renewcommand{\thetable}{S\arabic{table}}
\renewcommand{\thefigure}{S\arabic{figure}}
\renewcommand{\theHequation}{supp.\arabic{equation}}
\renewcommand{\theHtable}{supp.\arabic{table}}
\renewcommand{\theHfigure}{supp.\arabic{figure}}
\renewcommand{\theHsection}{supp.\Alph{section}}
\begin{center}
{\large\bfseries Supplemental Material}
\end{center}
In this Supplemental Material, we derive the results stated in the Letter and describe their numerical checks.
Sections~\ref{supp:A} and \ref{supp:B} fix the angular and radial conventions and the inner connection coefficient.
Sections~\ref{supp:C} and \ref{supp:D} connect it to physical QNMs through the boundary coefficient identity and the outer error estimate.
Section~\ref{supp:E} derives the oscillator representation and its asymptotic consequences, and describes the Argyres--Douglas limit.
Section~\ref{supp:F} separates physical frequency tests, boundary coefficient tests, and checks of the limiting equations.
References are shared with the Letter.
Throughout the Letter and this Supplemental Material, $f=O(g)$ means $|f|\le C|g|$ with a constant $C$ independent of the large parameter.

\section{Angular and radial equations}\label{supp:A}
\subsection{Angular convention}
With the conventions of Eq.~\eqref{eq:scales} and $u=\cos\theta$, the angular eigenvalue $A(c)$ is defined by
\begin{equation}
\frac{\dd}{\dd u}\left[(1-u^2)\frac{\dd S}{\dd u}\right]
+\left[c^2u^2-2csu+A(c)+s-\frac{(m+su)^2}{1-u^2}\right]S=0.
\label{supp:angular}
\end{equation}
The angular function $S(u)$ has the endpoint behaviors
\begin{equation}
S(u)\propto(1-u)^{|m+s|/2}\quad(u\to1),\qquad
S(u)\propto(1+u)^{|m-s|/2}\quad(u\to-1).
\label{supp:angularBC}
\end{equation}
We exclude the other endpoint branches, including logarithmic solutions when the indices coincide.
At $c=0$, $A(0)=\ell(\ell+1)-s(s+1)$, where $\ell\geq\max(|s|,|m|)$ is the spherical angular label.
The radial equations below contain the angular eigenvalue through
\begin{equation}
B=A(c)+s(s+1)-2mc+c^2.
\label{supp:Bdef}
\end{equation}

We continue the angular eigenvalue from its spherical value along the ray from $0$ to $c$.
For frequencies with positive real part, this ray lies slightly to the right of the negative imaginary axis.
The normal prolate branches satisfy
\begin{equation}
A(c)=\ii(2\bar L+1)c+O(1),
\label{supp:prolate}
\end{equation}
where $\bar L$ is the local oscillator index of the normal branch~\cite{BertiCardosoCasals,VickersCook}.
It need not equal $\ell$.
For the two families of the Letter, $(s,\ell,m)=(0,0,0)$ and $(-2,2,-2)$, we find $\bar L=0$.
The conjectured bound of Ref.~\cite{VickersCook} on the number of anomalous branches is zero for $s=0$ and for $s=-2$ with $|m|\ne2$.
It is one for $s=-2$ with $|m|=2$, and on our rays we find this branch at $\ell=3$.
Both physical and model roots use the eigenvalue at finite $c$.

\subsection{Two radial formulations}
The standard radial Teukolsky equation is~\cite{Teukolsky,Leaver}
\begin{equation}
\begin{gathered}
\Delta^{-s}\frac{\dd}{\dd r}\left(\Delta^{s+1}\frac{\dd R_{\rm T}}{\dd r}\right)
+\left[\frac{K_r^2-2\ii s(r-M)K_r}{\Delta}+4\ii s\omega r-B+s(s+1)\right]R_{\rm T}=0,\\
\Delta=r^2-2Mr+a^2,\qquad K_r=(r^2+a^2)\omega-am.
\end{gathered}
\label{supp:teuk}
\end{equation}
Here $r$ is the Boyer--Lindquist radius and $R_{\rm T}(r)$ is the standard radial function.
We write $\rho=\sqrt{1-\chi^2}$ and $r_\pm=M(1\pm\rho)$.
For $0\leq\chi<1$, the dimensionless coordinate and normal form wave are
\begin{equation}
z_{\rm T}=\frac{r-r_-}{2M\rho},\qquad
\Psi_{\rm T}=\Delta^{(s+1)/2}R_{\rm T}.
\label{supp:stdnormal}
\end{equation}
A direct substitution gives $\dd^2\Psi_{\rm T}/\dd r^2+Q_r\Psi_{\rm T}=0$, with
\begin{equation}
Q_r=\frac{K_r^2-2\ii s(r-M)K_r}{\Delta^2}
+\frac{4\ii s\omega r-B+s(s+1)}{\Delta}
-\frac{s+1}{2}\frac{\Delta''}{\Delta}
+\frac{1-s^2}{4}\left(\frac{\Delta'}{\Delta}\right)^2.
\label{supp:Qr}
\end{equation}
Primes in this equation denote $r$ derivatives.
The coefficient in the $z_{\rm T}$ normal form is $(2M\rho)^2Q_r$.
After this change of independent variable, we write the same normal form wave as $\Psi_{\rm T}(z_{\rm T})$.

The alternative radial equation used in the Letter is~\cite{HatsudaAlternative,Paper1}
\begin{equation}
\left[f\frac{\dd}{\dd z}f\frac{\dd}{\dd z}+\Omega^2-V(z)\right]\phi=0,
\qquad f=1-\frac1z,
\qquad
V=f\left[4c^2+\frac{4c(m-c)}z+\frac B{z^2}+\frac{1-s^2}{z^3}\right],
\label{supp:dual}
\end{equation}
where $B$ is given in Eq.~\eqref{supp:Bdef}.
We call it the dual equation below.
Its coordinate $z$ is not obtained by simply replacing $z_{\rm T}$ in the same radial wave function.
The two formulations are related through the confluent Heun parameter exchange discussed in Refs.~\cite{HatsudaAlternative,Paper1}.
The integral transformation of Whiting with an exponential kernel maps the standard radial equation to Eq.~\eqref{supp:dual}~\cite{Whiting}.
The boundary correspondence in the damped frequency domain requires analytic continuation.

For the dual equation, $\Psi=\sqrt f\,\phi$ satisfies $\Psi''+Q_z\Psi=0$, where primes now denote $z$ derivatives and
\begin{equation}
Q_z=\frac{\Omega^2z^4+(1-z)[4c^2z^3+4c(m-c)z^2+Bz-(s^2-1)]+z-3/4}
{z^2(1-z)^2}.
\label{supp:Qz}
\end{equation}
Before taking the scaling limit, the exact Laurent decomposition is
\begin{equation}
Q_z=\frac{1/4-s^2}{z^2}+\frac{B+1/2-s^2}{z}+Q_{\rm hol},\qquad
Q_{\rm hol}=\frac{\Omega^2z^2}{(1-z)^2}+\frac{B+4cm-s^2}{1-z}-4c^2
+\frac{3/4-z/2}{(1-z)^2}.
\label{supp:laurent}
\end{equation}
Here $Q_{\rm hol}$ is holomorphic at fixed parameters.
The term $\Omega^2z^2$ becomes leading only in the joint limit.

\subsection{Scaling of the inner equations}
We retain the convention of the Letter,
\begin{equation}
w_N=\frac{N+1/2}{2},\qquad \eps=w_N^{-1/4},\qquad
\chi=\xi\eps^3,\qquad \Omega=-\ii\eps^{-4}+O(1).
\label{supp:scales}
\end{equation}
The normal branch then has $c=-\ii\xi/(2\eps)+O(\eps^3)$ and $A=O(\eps^{-1})$.
Thus $c^2=O(\eps^{-2})$ is the leading part of $B$.
The balance in \eqrefp{supp:laurent} selects $z$ of order $\eps^2$.
The inner coordinates are $v=-\ii\sqrt\Omega\,z$ in the dual formulation and $v=-\ii\sqrt\Omega\,z_{\rm T}$ in the standard one.
The two limiting coordinates share the notation $v$ but differ at finite frequency.
We continue $\sqrt\Omega$ from $e^{-\ii\pi/4}\sqrt{w_N}$.

For the dual wave we set $y_{\rm d}(v)=\Psi(\ii v/\sqrt\Omega)$.
The transformed coefficient is $-Q_z/\Omega$.
If $\kappa_{\rm eff}=\ii B/\sqrt\Omega$ is retained without expanding $A(c)$, its first terms are
\begin{equation}
-\frac{Q_z}{\Omega}=v^2+\frac{\kappa_{\rm eff}}v+\frac{1/4-s^2}{v^2}
+\eps^2\left[2e^{3\pi\ii/4}v^3+\frac{e^{3\pi\ii/4}(1/2-s^2)}v
+e^{3\pi\ii/4}\kappa_{\rm eff}-\ii\xi^2\right]+O(\eps^3).
\label{supp:dualexp}
\end{equation}
This expansion holds at fixed $v\ne0$ and is nonuniform at infinity.
The extra $1/v$ coefficient $\ii(1/2-s^2)/\sqrt\Omega$ starts at order $\eps^2$ and is excluded from $\kappa_{\rm eff}$.
Using \eqrefp{supp:prolate}, we obtain
\begin{equation}
\kappa_{\rm eff}=e^{-\ii\pi/4}
\left[\frac{\xi^2}{4}-\frac{\eps\xi}{2}(2\bar L+1+2\ii m)+O(\eps^2)\right],
\qquad \kappa_0=\frac{e^{-\ii\pi/4}}4\xi^2.
\label{supp:kappalimit}
\end{equation}

For the standard wave we set $y_{\rm T}(v)=\Psi_{\rm T}(\ii v/\sqrt\Omega)$.
In \eqrefp{supp:Qr}, the term quadratic in $K_r$ gives $v^2$ in the limit.
The spin term proportional to $(r-M)K_r$ gives $-\ii s$.
The term $[4\ii s\omega r-B+s(s+1)]/\Delta$ gives $4\ii s+\kappa_0/v$.
The $\Delta''$ term vanishes in the limit, while the last term gives $(1-s^2)/(4v^2)$.
Consequently, the two limiting ordinary differential equations are
\begin{align}
y_{\rm T}''+\left[v^2+3\ii s+\frac{\kappa_0}{v}+\frac{1-s^2}{4v^2}\right]y_{\rm T}&=0,
\label{supp:innerT}\\
y_{\rm d}''+\left[v^2+\frac{\kappa_0}{v}+\frac{1/4-s^2}{v^2}\right]y_{\rm d}&=0.
\label{supp:innerD}
\end{align}
Here and below a prime on an inner wave denotes a $v$ derivative.
For fixed angular labels, the separate angular regime with $|c|$ of order one corresponds to $\xi$ of order $\eps$.
The known spherical to prolate behavior~\cite{BertiCardosoCasals,VickersCook} therefore does not occur on the radial scale where $\xi$ is of order one.

\section{Connection coefficients and Darboux transformation}\label{supp:B}
\subsection{The auxiliary sextic problems}
We now replace $\kappa_0$ by a general complex parameter $\kappa$ in Eqs.~\eqref{supp:innerT} and \eqref{supp:innerD}.
The connection coefficients are functions of this parameter.
We use
\begin{equation}
\gamma=\frac{e^{\ii\pi/4}}{\sqrt2},\qquad v=\gamma x^2,
\qquad \varphi(x)=\left(\frac{\dd v}{\dd x}\right)^{-1/2}y(v(x)),
\qquad E=4\gamma\kappa=2\sqrt2e^{\ii\pi/4}\kappa.
\label{supp:liouville}
\end{equation}
This $E$ is the energy of the sextic oscillator in Eq.~\eqref{eq:sextic} of the Letter.
The coordinate $x$ is also auxiliary.
For a general inner coefficient $Q_v$, the transformed equation has coefficient
$(v')^2Q_v+\{v,x\}/2$, where $\{v,x\}=v'''/v'-3(v''/v')^2/2$ is the Schwarzian derivative.
For $v=\gamma x^2$, it is $\{v,x\}=-3/(2x^2)$.
We find
\begin{align}
H_{\rm T}\varphi_{\rm T}=E\varphi_{\rm T},\qquad
H_{\rm T}&=-\partial_x^2+x^6+6sx^2+\frac{s^2-1/4}{x^2},\label{supp:HT}\\
H_{\rm d}\varphi_{\rm d}=E\varphi_{\rm d},\qquad
H_{\rm d}&=-\partial_x^2+x^6+\frac{4s^2-1/4}{x^2}.\label{supp:HD}
\end{align}
Both belong to the family
\begin{equation}
H(\alpha,l)=-\partial_x^2+x^6+\alpha x^2+\frac{l(l+1)}{x^2}
\label{supp:Hgeneral}
\end{equation}
studied by Dorey, Dunning, and Tateo~\cite{DoreyDunningTateo}.
Their parameters are $(\alpha,l)=(6s,|s|-1/2)$ for the standard equation and $(0,2|s|-1/2)$ for the dual equation.
The auxiliary index $l$ differs from the angular label $\ell$.

The constant $3\ii s$ in \eqrefp{supp:innerT} changes the powers of the asymptotic waves.
The unit amplitude conventions are
\begin{equation}
y^{\rm T}_{\pm}\sim v^{-1/2\mp3s/2}e^{\pm\ii v^2/2},\qquad
y^{\rm d}_{\pm}\sim v^{-1/2}e^{\pm\ii v^2/2}.
\label{supp:waves}
\end{equation}
For the general problem \eqrefp{supp:Hgeneral}, the corresponding powers are $v^{-1/2\mp\alpha/4}$.
All fractional powers are defined on the continuously lifted argument of the continuation path.

\subsection{Sectorial solutions and normalization}
We use the normalization of Eqs.~(2.2) and (2.3) of Ref.~\cite{DoreyDunningTateo}.
The canonical solution on the positive $x$ axis is
\begin{equation}
Y(x,E,\alpha,l)\sim\frac{x^{-3/2-\alpha/2}}{\sqrt{2\ii}}e^{-x^4/4},
\qquad \sqrt{2\ii}=\sqrt2e^{\ii\pi/4}.
\label{supp:Y}
\end{equation}
It is the subdominant solution in the sector $|\arg x|<\pi/8$.
We define $q=e^{\ii\pi/4}$ and its rotated solutions by
\begin{equation}
Y_k(x,E,\alpha,l)=q^{k(1+\alpha)/2}
Y(q^{-k}x,\ii^kE,(-1)^k\alpha,l),\qquad
\mathcal S_k=\left\{x:\left|\arg x-\frac{k\pi}{4}\right|<\frac\pi8\right\}.
\label{supp:rotated}
\end{equation}
Each $Y_k$ solves the equation at $(E,\alpha,l)$ and decays in $\mathcal S_k$.
An odd rotation reverses $\alpha$.
With $W_x[f,g]=fg'-f'g$, the prefactors give $W_x[Y_{-1},Y_0]=W_x[Y_0,Y_1]=1$.

The inner continuation of the Letter starts on the ray $\arg v=0^+$, which we call the A side.
After a clockwise rotation by $3\pi/2$, it ends on the ray $\arg v=-3\pi/2+0^+$, the B side.
With $\log v=\log\gamma+2\log x$, these become
\begin{equation}
\arg x_A=-\frac\pi8+0^+,\qquad
\arg x_B=-\frac{7\pi}{8}+0^+.
\label{supp:sectorsAB}
\end{equation}
We denote by $y_A$ the wave with the upper sign on the A side, by $y_A^{\rm cont}$ its continuation to the B side, and by $y_B^-$ the wave with the lower sign on the B side.
Thus $y_A$ is the boundary value of $Y_0$, and $y_B^-$ is the boundary value of $Y_{-3}$.
For either value of $\alpha$, the normalization factors are
\begin{equation}
\begin{aligned}
y_A&=\mathcal N_A\sqrt{2\gamma x}\,Y_0,
&\mathcal N_A&=e^{\ii\pi/4}\gamma^{-1-\alpha/4},\\
y_B^-&=\mathcal N_B\sqrt{2\gamma x}\,Y_{-3},
&\mathcal N_B&=e^{-\ii\pi/4}\gamma^{-1+\alpha/4}.
\end{aligned}
\label{supp:normalization}
\end{equation}
For example, on the lifted B sector,
$Y_{-3}\sim\ii x^{-3/2+\alpha/2}e^{x^4/4}/\sqrt{2\ii}$.
This fixes the second factor in \eqrefp{supp:normalization}, including its phase.
The Liouville transform preserves the Wronskian, and $\mathcal N_A\mathcal N_B=\gamma^{-2}=-2\ii$.
Hence both connection coefficients have the same representation,
\begin{equation}
T_s(\kappa)=\frac1{2\ii}W_v[y_A^{\rm cont},y_B^-]
=-W_x[Y_0,Y_{-3}].
\label{supp:Tbridge}
\end{equation}
All waves are on the same lifted sheet.
Only canonical decaying solutions enter, so the definition is independent of adding a decaying wave to a dominant one.
It fixes the dominant B coefficient to be $-T_s$.

\subsection{Fusion with the deformation}
The basic Stokes multiplier is denoted by $C(E,\alpha,l)$ here.
It is called $T(E,\alpha,l)$ in Ref.~\cite{DoreyDunningTateo}, but differs from our fused coefficient $T_s$.
Its defining relation is
\begin{equation}
C(E,\alpha,l)Y_0=Y_{-1}+Y_1.
\label{supp:basicC}
\end{equation}
Replacing $x$, $E$, $\alpha$ by $q^{-k}x$, $\ii^kE$, $(-1)^k\alpha$ in \eqrefp{supp:basicC} and using \eqrefp{supp:rotated}, we obtain, for every integer $k$,
\begin{equation}
\begin{aligned}
C(\ii^kE,\alpha,l)\,Y_k&=Y_{k-1}+Y_{k+1}, &&k\ \text{even},\\
C(\ii^kE,-\alpha,l)\,Y_k&=q^{\alpha}Y_{k-1}+q^{-\alpha}Y_{k+1}, &&k\ \text{odd}.
\end{aligned}
\label{supp:generalk}
\end{equation}
The factors $q^{\pm\alpha}$ appear only for odd $k$, where the rotation reverses $\alpha$.
The cases $k=-1$ and $k=-2$ give
\begin{equation}
\begin{aligned}
Y_{-2}&=q^{-\alpha}C(-\ii E,-\alpha,l)Y_{-1}-q^{-2\alpha}Y_0,\\
Y_{-3}&=C(-E,\alpha,l)Y_{-2}-Y_{-1}.
\end{aligned}
\label{supp:twofusion}
\end{equation}
Eliminating $Y_{-2}$, we obtain
\begin{equation}
Y_{-3}=\left[q^{-\alpha}C(-E,\alpha,l)C(-\ii E,-\alpha,l)-1\right]Y_{-1}
-q^{-2\alpha}C(-E,\alpha,l)Y_0.
\label{supp:fusedbasis}
\end{equation}
Since $W_x[Y_{-1},Y_0]=1$, the coefficient of $Y_{-1}$ equals $-W_x[Y_0,Y_{-3}]$.
Thus
\begin{align}
T_s^{\rm T}(\kappa)+1
&=e^{-3\pi\ii s/2}
C(-E,6s,|s|-1/2)C(-\ii E,-6s,|s|-1/2),\label{supp:stdFusion}\\
T_s^{\rm d}(\kappa)+1&=C_{\rm d}(-E)C_{\rm d}(-\ii E),
\qquad C_{\rm d}(E)=C(E,0,2|s|-1/2).\label{supp:dualFusion}
\end{align}
The standard prefactor is $1$, $-\ii$, and $-1$ for $s=0,-1,-2$, respectively.
The common constant $1/\sqrt{2\ii}$ in \eqrefp{supp:Y} does not change $C_{\rm d}$ relative to the unit amplitude convention of the Letter.

\subsection{Equality of the connection coefficients}
We prove the equality for $s=-J$, with $J=1,2$.
The scalar case $J=0$ is immediate because the two limiting equations coincide.
The standard parameters obey the quasi-exact solvability condition
$\alpha=-(4J+2l+1)$ with $l=J-1/2$.
The transformation of Eqs.~(6.13)--(6.17) of Ref.~\cite{DoreyDunningTateo} gives
\begin{equation}
\widehat\alpha=2J-2l-1=0,\qquad
\widehat l=J+l=2J-\frac12.
\label{supp:QESmap}
\end{equation}
These are precisely the dual parameters.
The differential intertwining operator of their Eq.~(8.11) becomes
\begin{equation}
\mathcal Q_J=e^{-x^4/4}x^{2J+1/2}
\left(\frac1x\frac{\dd}{\dd x}\right)^J
 e^{x^4/4}x^{-J-1/2},\qquad
\mathcal Q_JH_{\rm T}=H_{\rm d}\mathcal Q_J.
\label{supp:intertwiner}
\end{equation}
The rightmost multiplication acts first.
Explicitly,
\begin{equation}
\mathcal Q_1=\partial_x+x^3-\frac{3}{2x},\qquad
\mathcal Q_2=\partial_x^2+\left(2x^3-\frac6x\right)\partial_x
+x^6-3x^2+\frac{45}{4x^2}.
\label{supp:Q12}
\end{equation}
The coefficients are single valued away from $x=0$, so these operators commute with continuation along our lifted path.

For odd $k$, \eqrefp{supp:rotated} gives
\begin{equation}
Y_k^{\rm T}\sim\frac{\ii^k}{\sqrt{2\ii}}x^{-3/2-3J}e^{x^4/4},\qquad
Y_k^{\rm d}\sim\frac{\ii^k}{\sqrt{2\ii}}x^{-3/2}e^{x^4/4}.
\label{supp:oddwaves}
\end{equation}
Acting with $\mathcal Q_J$ on the first expression produces the factor $2^J$.
Its image solves the dual equation and is subdominant in the same sector.
The uniqueness of the normalized subdominant solution therefore implies the exact relations
\begin{equation}
\mathcal Q_JY_{-1}^{\rm T}=2^JY_{-1}^{\rm d},\qquad
\mathcal Q_JY_{-3}^{\rm T}=2^JY_{-3}^{\rm d}.
\label{supp:oddaction}
\end{equation}
The same normalized-solution argument gives
\begin{equation}
\mathcal Q_JY_0^{\rm T}=2^{-J}P_J(E)Y_0^{\rm d},\qquad
P_1(E)=E,\qquad P_2(E)=E^2-48.
\label{supp:evenaction}
\end{equation}
The normalization can also be checked from
$W_x[\mathcal Q_Jf,\mathcal Q_Jg]=P_J(E)W_x[f,g]$ for two solutions of $H_{\rm T}f=Ef$.
Both identities follow directly from \eqrefp{supp:Q12} and the differential equation.

We write \eqrefp{supp:fusedbasis} in the standard problem as
\begin{equation}
Y_{-3}^{\rm T}=T_s^{\rm T}(\kappa)Y_{-1}^{\rm T}
+b_{\rm T}(E)Y_0^{\rm T},
\label{supp:expandT}
\end{equation}
where $b_{\rm T}(E)$ is the other expansion coefficient.
Applying $\mathcal Q_J$ and dividing by $2^J$ gives
\begin{equation}
Y_{-3}^{\rm d}=T_s^{\rm T}(\kappa)Y_{-1}^{\rm d}
+\frac{P_J(E)}{4^J}b_{\rm T}(E)Y_0^{\rm d}.
\label{supp:expandD}
\end{equation}
The pair $(Y_{-1}^{\rm d},Y_0^{\rm d})$ is a basis for every $E$, since its Wronskian is one.
The uniqueness of its expansion coefficients proves
\begin{equation}
T_s^{\rm T}(\kappa)=T_s^{\rm d}(\kappa),\qquad s=0,-1,-2.
\label{supp:Tidentity}
\end{equation}
We never divided by $P_J(E)$.
The argument therefore also applies at $E=0$ for $J=1$ and at $E=\pm4\sqrt3$ for $J=2$.
At these quasi-exactly solvable energies the even solution is annihilated, but the odd images in \eqrefp{supp:oddaction} remain nonzero.

The spectra of the two half-line problems differ by these $J$ exceptional levels~\cite{DoreyDunningTateo}, and the basic multipliers differ as well.
The fused coefficient \eqref{supp:Tbridge} is the same for both problems.
The argument applies the rotated solutions and the intertwiner of Ref.~\cite{DoreyDunningTateo} to this coefficient.

\section{The standard and dual boundary conditions}\label{supp:C}
The result of Sec.~\ref{supp:B} concerns two limiting differential equations.
We now relate the physical boundary conditions at finite frequency.
This step is separate from the outer approximation in Sec.~\ref{supp:D}.
We use the confluent Heun representation of Ref.~\cite{Paper1} and the spectral symmetry studied in Ref.~\cite{CasalsTeixeira}.
The integral transform below is related to the Euler symmetries of confluent Heun equations~\cite{KazakovSlavyanov}.
Integral transformations of this type, with Euler and Laplace kernels, were also used by Whiting to prove mode stability~\cite{Whiting}.
We need a relation between \emph{incoming coefficients}, rather than an equality between the wave functions at the same radial coordinate.
A similar relation for the reduced and doubly confluent Heun equations is proved with a Laplace-type transform in Ref.~\cite{GiorgiGrassi}.

\subsection{Boundary waves and the Jaff\'e recurrence}
For three parameters $\mu_1$, $\mu_2$, $\mu_3$ and a permutation $(i,j,k)$ of $(1,2,3)$, we write
\begin{equation}
\begin{split}
0=\bigg[&z(z-1)\partial_z^2-p^2z(z-1)-\mu_kp(2z-1)
+\mathcal E+\frac14\\
&-\frac{\mu_i\mu_j}{z}-\frac{(\mu_i+\mu_j)^2-1}{4z(z-1)}\bigg]\psi.
\end{split}
\label{supp:CHE}
\end{equation}
The standard and dual radial equations correspond to $(i,j,k)=(1,2,3)$ and $(1,3,2)$, respectively, with
\begin{equation}
\begin{gathered}
\mu_1=s+\ii\Omega,\quad
\mu_2=\frac{\ii(\Omega-\chi m)}{\rho},\quad
\mu_3=-s+\ii\Omega,\\
p=\ii\rho\Omega,\qquad
\mathcal E=-A(c)-s(s+1)+2\Omega^2-c^2-\frac14.
\end{gathered}
\label{supp:CHEmap}
\end{equation}
The standard coordinate is $z_{\rm T}=(r-r_-)/(2M\rho)$.
The solution $\psi$ is the normal form wave $\Psi_{\rm T}$ of Eq.~\eqref{supp:stdnormal} in the standard case and $\Psi=\sqrt f\,\phi$ in the dual case.
These statements follow by substitution in Eq.~\eqref{supp:teuk} and Eq.~\eqref{supp:dual}.

We normalize the horizon solution by
$H_{ijk}=(z-1)^{(1-\mu_i-\mu_j)/2}[1+O(z-1)]$.
The outgoing solution $O_{ijk}$ is first specified in a decaying sector by
$O_{ijk}\sim e^{pz}z^{\mu_k}$ at infinity.
It is then continued in the frequency through the \emph{upper} half of the $p$ plane.
This is our retarded prescription.
For the incoming coefficient we use the Wronskian definition
\begin{equation}
\mathcal A_{\rm in}^{ijk}=\frac{W_z[H_{ijk},O_{ijk}]}{2p}.
\label{supp:incomingCHE}
\end{equation}
Indeed, the complementary incoming wave $I_{ijk}$ has unit asymptotic amplitude
$e^{-pz}z^{-\mu_k}$ and $W_z[O_{ijk},I_{ijk}]=-2p$.
The zero of \eqrefp{supp:incomingCHE} is the quasinormal-mode condition on the specified sheet.

With $\eta=(z-1)/z$, the horizon solution is
\begin{equation}
H_{ijk}=e^{p(z-1)}z^{(\mu_i+\mu_j+2\mu_k-1)/2}
(z-1)^{-(\mu_i+\mu_j-1)/2}\sum_{n\ge0}b_n\eta^n,
\quad b_0=1.
\label{supp:CHEJaffe}
\end{equation}
Its three-term recurrence is
$\alpha_n b_{n+1}+\beta_n b_n+\gamma_n b_{n-1}=0$, where
\begin{equation}
\begin{aligned}
\alpha_n&=(n+1)(n+1-\mu_i-\mu_j),\\
\beta_n&=\mathcal E+\frac14+2n(p-n)
+(2n+1-p)(\mu_1+\mu_2+\mu_3-1)\\
&\hspace{1.5cm}-(\mu_1\mu_2+\mu_1\mu_3+\mu_2\mu_3),\\
\gamma_n&=(n-\mu_i-\mu_k)(n-\mu_j-\mu_k).
\end{aligned}
\label{supp:CHErec}
\end{equation}
The diagonal coefficient $\beta_n$ and the adjacent product
\begin{equation}
\alpha_{n-1}\gamma_n=n(n-\mu_1-\mu_2)(n-\mu_1-\mu_3)(n-\mu_2-\mu_3)
\label{supp:adjacent}
\end{equation}
are the same for all permutations.
This is the origin of the common characteristic continued fraction.
In fact, when $1-\mu_i-\mu_j$ and $1-\mu_i-\mu_k$ are not nonpositive integers,
\begin{equation}
\widetilde b_n=\frac{(1-\mu_i-\mu_j)_n}{(1-\mu_i-\mu_k)_n}\,b_n,
\qquad (i,j,k)\longmapsto(i,k,j),
\label{supp:Pochhammer}
\end{equation}
intertwines the two recurrences.
Here $(\sigma)_n=\Gamma(\sigma+n)/\Gamma(\sigma)$ is the Pochhammer symbol.
The multiplier is nonzero at every $n$.
It also preserves the ratio of any minimal solution to any independent dominant solution.
This statement alone does not yet fix the outgoing wave in the damped frequency half plane.

\subsection{Incoming amplitudes and analytic continuation}
The normalization can be obtained without relying on the zeros of a continued fraction.
In this subsection we treat $\mu_1$, $\mu_2$, $\mu_3$, $p$, and $\mathcal E$ as independent complex parameters.
For $\Re(1-\mu_i-\mu_j)>0$ and $\Re(\mu_j-\mu_k)>0$, the Euler beta integral gives
\begin{equation}
\widetilde b(\eta)=
\frac{\Gamma(1-\mu_i-\mu_k)}{\Gamma(1-\mu_i-\mu_j)\Gamma(\mu_j-\mu_k)}
\int_0^1t^{-\mu_i-\mu_j}(1-t)^{\mu_j-\mu_k-1}b(\eta t)\,\dd t,
\label{supp:EulerCHE}
\end{equation}
where $b(\eta)=\sum_{n\ge0}b_n\eta^n$ and
$\widetilde b(\eta)=\sum_{n\ge0}\widetilde b_n\eta^n$.
The identity holds near the horizon by convergence of the two series.
We first work with $\Re p<0$, so the incoming wave increases along positive real $z$.
By Eq.~\eqref{supp:CHEJaffe}, the incoming contribution to $b$ is
\begin{equation}
b(\eta)=e^p\mathcal A_{\rm in}^{ijk}
 e^{-2pz}z^{-2\mu_k}[1+O(z^{-1})]+\text{nonincreasing terms}.
\label{supp:incomingAsym}
\end{equation}
In \eqrefp{supp:EulerCHE} this contribution is governed by $t=1$.
Putting $t=1-\tau/z^2$, we obtain
\begin{equation}
\begin{aligned}
\int_0^1 &t^{-\mu_i-\mu_j}(1-t)^{\mu_j-\mu_k-1}b(\eta t)\,\dd t\\
&\sim e^p\mathcal A_{\rm in}^{ijk}
 e^{-2pz}z^{-2\mu_j}
 \int_0^\infty \tau^{\mu_j-\mu_k-1}e^{2p\tau}\,\dd\tau.
\end{aligned}
\label{supp:EulerEndpoint}
\end{equation}
The remaining integral is $\Gamma(\mu_j-\mu_k)(-2p)^{\mu_k-\mu_j}$.
Matching the coefficient of the growing wave gives
\begin{equation}
\mathcal A_{\rm in}^{ikj}=
\frac{\Gamma(1-\mu_i-\mu_k)}{\Gamma(1-\mu_i-\mu_j)}
(-2p)^{\mu_k-\mu_j}\mathcal A_{\rm in}^{ijk}.
\label{supp:CHEidentity}
\end{equation}
This calculation is first valid in the open set where the Euler integral converges and the incoming exponential dominates.
The physical values \eqref{supp:CHEmap} lie outside this set and are reached by analytic continuation in the parameters.
It is an identity between connection coefficients, including at their zeros by analytic dependence.

For the continuation we choose $\Im p>0$ and
$-\pi<\arg(-2p)<0$.
In this domain, the outgoing solution is fixed by its unit amplitude along a ray on which it decays.
Overlapping decaying sectors give the same solution, so it is independent of the ray.
By the Volterra equation at infinity, it is holomorphic in the parameters away from the Frobenius resonances.
This construction supplies the analytic continuation of Eq.~\eqref{supp:CHEidentity} to the QNM region with $\Re\Omega>0$.
Merely imposing an increasing leading exponential on the positive real radial axis would not select that solution.
The spectral symmetry in Ref.~\cite{CasalsTeixeira} was proved for the mode-stability half plane.
Here we also specify the outgoing analytic continuation needed in the damped region.

For $(i,j,k)=(1,2,3)$, Eq.~\eqref{supp:CHEidentity} reads
\begin{equation}
\mathcal A_{\rm in}^{\rm d}=G(\Omega,\chi,s,m)
\mathcal A_{\rm in}^{\rm T},\qquad
G=\frac{\Gamma(1-\mu_1-\mu_3)}
{\Gamma(1-\mu_1-\mu_2)}
(-2p)^{\mu_3-\mu_2}.
\label{supp:Gfactor}
\end{equation}
Here $\mathcal A_{\rm in}^{\rm T}=\mathcal A_{\rm in}^{123}$ and $\mathcal A_{\rm in}^{\rm d}=\mathcal A_{\rm in}^{132}$ are the incoming coefficients of the standard and dual equations.
Gamma functions have no zeros.
Their poles are at the nonpositive integers.
In particular, when $\Omega=d_R-\ii w$ with $d_R>0$ and
$(1+\rho)d_R>\chi m$, both gamma arguments are nonreal.
Then $G$ and its inverse are locally holomorphic and nonzero.
These conditions hold for sufficiently large $w$ in the compact positive-real-part strip of Sec.~\ref{supp:D}, with fixed $m$ and $\chi=O(w^{-3/4})$.
The wave-function conversions back to $R_{\rm T}$ and $\phi$ introduce only nonzero factors.
Thus
\begin{equation}
\mathcal A_{\rm in}^{\rm T}=0
\quad\Longleftrightarrow\quad
\mathcal A_{\rm in}^{\rm d}=0
\label{supp:QNMzeroEquality}
\end{equation}
on the common nonresonant, retarded continuation sheet.
The standard and dual equations therefore have the same QNM frequencies in the region used in Sec.~\ref{supp:D}.
Reference~\cite{Paper1} assumed this in the lower half plane.

\section{Outer matching and the QNM condition}\label{supp:D}
We derive the matching estimate for the dual equation and then apply the boundary correspondence of Sec.~\ref{supp:C}.
We give the essential steps of the estimates.
We consider frequencies and angular input satisfying
\begin{equation}
\begin{gathered}
\Omega=d_R-\ii w,\quad 0<d_0\le d_R\le d_1,\quad w\to\infty,\\
0\le\chi\le C_\chi w^{-3/4},\qquad |B|\le C_B\sqrt w,
\end{gathered}
\label{supp:P1domain}
\end{equation}
where $s,m$ are fixed and all constants are independent of $w$.
On this domain $c=\chi\Omega/2=O(w^{1/4})$ and $\rho^{-1}=1+O(w^{-3/2})$.
The matching uses the angular eigenvalue only through this bound on $B$.
The assumptions on the angular branch enter only in the last subsection.

\subsection{Boundary condition and monodromy}
We write the dual equation in the coordinate $\widehat x=z+\log(1-z)$, which satisfies $\dd\widehat x/\dd z=1/f$,
\begin{equation}
\phi_{\widehat x\widehat x}+[\Omega^2-V(z)]\phi=0,
\qquad f=1-z^{-1},
\label{supp:Schrodinger}
\end{equation}
where $V$ is given in Eq.~\eqref{eq:radial}.
We normalize the horizon and infinity waves by
\begin{equation}
\begin{aligned}
H_{\rm in/out}&\sim(z-1)^{\mp\ii\Omega},\\
F_{\rm out/in}&\sim e^{\pm\ii\rho\Omega z}z^{\pm\ii\alpha_\infty},
\qquad \alpha_\infty=\frac{\Omega-\chi m}{\rho}.
\end{aligned}
\label{supp:boundary}
\end{equation}
We also write $\beta=\alpha_\infty-\Omega$, which is $O(w^{-1/2})$ on the domain \eqref{supp:P1domain}.
The outgoing infinity wave is the canonical solution selected in a decaying sector and continued through the right side of the frequency plane.
The physical standard Teukolsky boundary waves are
\begin{equation}
R_{\rm H}^{\rm in}\sim(r-r_+)^{-s-\ii\sigma_+},\quad
R_\infty^{\rm out}\sim e^{\ii\omega r}r^{-1-2s+\ii\Omega},\quad
\sigma_+=\frac{(1+\rho)\Omega-\chi m}{2\rho}.
\label{supp:standardBC}
\end{equation}
Their boundary-value zeros agree with those of \eqrefp{supp:boundary} under the assumptions in Sec.~\ref{supp:C}.

We define the connection matrix by
\begin{equation}
(H_{\rm in},H_{\rm out})=(F_{\rm out},F_{\rm in})S,
\quad S=\begin{pmatrix}A_{\rm out}&B_{\rm out}\\A_{\rm in}&B_{\rm in}\end{pmatrix}.
\label{supp:connectionmatrix}
\end{equation}
With the Wronskian $W_{\widehat x}[\phi_1,\phi_2]=\phi_1\phi_{2,\widehat x}-\phi_{1,\widehat x}\phi_2$, the normalized boundary limits give
\begin{equation}
W_{\widehat x}[H_{\rm in},H_{\rm out}]=2\ii\Omega,\quad
W_{\widehat x}[F_{\rm out},F_{\rm in}]=-2\ii\rho\Omega,
\quad\det S=-\rho^{-1}.
\label{supp:boundaryWronskians}
\end{equation}
Counterclockwise continuation around the outer horizon gives the monodromy matrix
\begin{equation}
\mathsf M=S\operatorname{diag}(e^{2\pi\Omega},e^{-2\pi\Omega})S^{-1}.
\label{supp:Mmatrix}
\end{equation}
Its outgoing projection defines the quantization function and satisfies the exact identity
\begin{equation}
\mathcal D(\Omega)=e^{4\pi\Omega}-e^{2\pi\Omega}\mathsf M_{11}
=\rho(e^{4\pi\Omega}-1)A_{\rm in}B_{\rm out}.
\label{supp:exactmonodromy}
\end{equation}
The entry can also be computed directly from a continued outgoing wave,
\begin{equation}
\mathsf M_{11}=
\frac{W_{\widehat x}[\mathcal M_1 F_{\rm out},F_{\rm in}]}
{W_{\widehat x}[F_{\rm out},F_{\rm in}]}.
\label{supp:projection}
\end{equation}
Here $\mathcal M_1$ denotes continuation along the counterclockwise loop about the horizon specified in Sec.~\ref{supp:Dinner}.
The numerator requires both solutions on the same sheet.
Since $A_{\rm in}$ and $B_{\rm out}$ are Wronskians with the infinity waves, $\mathcal D$ is written with Wronskians as
\begin{equation}
\mathcal D(\Omega)=e^{4\pi\Omega}
+\frac{e^{2\pi\Omega}}{2\ii\rho\Omega}
W_{\widehat x}[\mathcal M_1F_{\rm out},F_{\rm in}]
=\frac{e^{4\pi\Omega}-1}{4\rho\Omega^2}\,
W_{\widehat x}[H_{\rm in},F_{\rm out}]\,W_{\widehat x}[H_{\rm out},F_{\rm in}].
\label{supp:factorization}
\end{equation}
The first Wronskian on the right vanishes exactly at the QNM frequencies of the dual equation.
The second one is $-2\ii\rho\Omega B_{\rm out}$, and it is nonzero for large $w$.
If $B_{\rm out}$ vanished, $H_{\rm out}$ would be proportional to $F_{\rm in}$ and would decay at both ends of the real exterior $z>1$.
Multiplying \eqrefp{supp:Schrodinger} by the complex conjugate wave and integrating would then give $d_R^2-w^2\ge-\sup_{z>1}|V|$.
This is impossible for large $w$, since $\sup_{z>1}|V|=O(\sqrt w)$ on the domain of \eqrefp{supp:P1domain}.
Since also $e^{4\pi\Omega}\ne1$, the zeros of $\mathcal D$ are precisely the QNM frequencies of the dual equation.
By Eq.~\eqref{supp:QNMzeroEquality}, they are also the standard Teukolsky QNM frequencies on the same retarded sheet.

\subsection{Inner connection and matching}\label{supp:Dinner}
We use a coordinate in which the rapid phase is exact,
\begin{equation}
\zeta^2=2\Omega\widehat x,\qquad
\mathscr U=e^{\pi\Omega}\zeta^{-1/2}\phi.
\label{supp:exactphase}
\end{equation}
All logarithms and square roots are continued along the path specified below, and $\zeta\simeq-\ii\sqrt\Omega\,z$ near $z=0$.
The function $\mathscr U$ satisfies
\begin{equation}
\mathscr U_{\zeta\zeta}+
\left[\zeta^2-\frac{\zeta^2}{\Omega^2}V(z)
-\frac{3}{4\zeta^2}\right]\mathscr U=0,
\qquad
W_{\widehat x}[\phi_1,\phi_2]=\Omega e^{-2\pi\Omega}W_\zeta[\mathscr U_1,\mathscr U_2].
\label{supp:exactLiouville}
\end{equation}
Since $\zeta/\sqrt\Omega$ is a function of $z$ alone with a simple zero at $z=0$, the map is invertible for $|\zeta|\le L$, where $L$ is a small fixed multiple of $\sqrt w$.
Near $z=0$, the coefficient $-\zeta^2V/\Omega^2$ in \eqrefp{supp:exactLiouville} has the leading terms $\kappa_{\rm eff}/\zeta+(1-s^2)/\zeta^2$ and a small constant.
We therefore approximate \eqrefp{supp:exactLiouville} by the auxiliary equation
\begin{equation}
y_{\zeta\zeta}+\left[\zeta^2+2\beta+\frac{\kappa_{\rm eff}}{\zeta}
+\frac{1/4-s^2}{\zeta^2}\right]y=0,
\qquad\kappa_{\rm eff}=\frac{\ii B}{\sqrt\Omega}.
\label{supp:innerBeta}
\end{equation}
At $\beta=0$, it is the inner equation \eqref{eq:inner} of the Letter with $\kappa_{\rm eff}$ in place of $\kappa_0$.
To measure the error of this approximation, we write the exact equation \eqrefp{supp:exactLiouville} in the same form,
\begin{equation}
\mathscr U_{\zeta\zeta}+\left[\zeta^2+2\beta+\frac{\kappa_{\rm eff}}{\zeta}
+\frac{1/4-s^2}{\zeta^2}+g_\beta(\zeta)\right]\mathscr U=0,
\qquad
g_\beta=-\frac{\zeta^2}{\Omega^2}V(z)-2\beta-\frac{\kappa_{\rm eff}}{\zeta}-\frac{1-s^2}{\zeta^2}.
\label{supp:deviation}
\end{equation}
The function $g_\beta$ is the deviation of the exact equation from the auxiliary equation.
The matching estimate rests on its smallness along the inner path.
The constant $2\beta$ is of order $w^{-1/2}$.
It equals the leading constant part $4c(c-m)/\Omega$ of $-\zeta^2V/\Omega^2$ up to $O(w^{-2})$, by $c=\chi\Omega/2$ and $\rho^2=1-\chi^2$.
Without this term, $g_\beta$ would contain a constant of order $w^{-1/2}$, which would produce a logarithm of $w$ in the integral of \eqrefp{supp:innerIntegral} below.
Keeping $2\beta$ is analogous to using Coulomb waves instead of plane waves in scattering theory.
The term proportional to $B$ needs a separate check, since $B$ can be of order $\sqrt w$.
With $t=\zeta/(-\ii\sqrt\Omega)=z+z^2/3+O(z^3)$, this term of $-\zeta^2V/\Omega^2$ is exactly $-B(1-z)(t/z)^3/(\Omega t)$.
Since $(1-z)(t/z)^3=1-z^2/12+O(z^3)$ and $-B/(\Omega t)=\kappa_{\rm eff}/\zeta$, it equals $\kappa_{\rm eff}/\zeta$ up to $O(|\zeta|/w)$.
Thus it leaves no constant of order $B/\Omega$ in $g_\beta$.
For $|\zeta|\le L$, the point $z$ stays in a fixed small neighborhood of $z=0$.
Expanding $g_\beta$ there and using $|B|\le C_B\sqrt w$, we find for a fixed $L_0>0$
\begin{equation}
|g_\beta(\zeta)|\le C\left(\frac{|\zeta|}{w}+\frac{w^{-1/2}}{|\zeta|}\right)
\qquad (L_0\le|\zeta|\le L).
\label{supp:gbound}
\end{equation}

\begin{figure}[t]
\centering
\includegraphics[width=\textwidth]{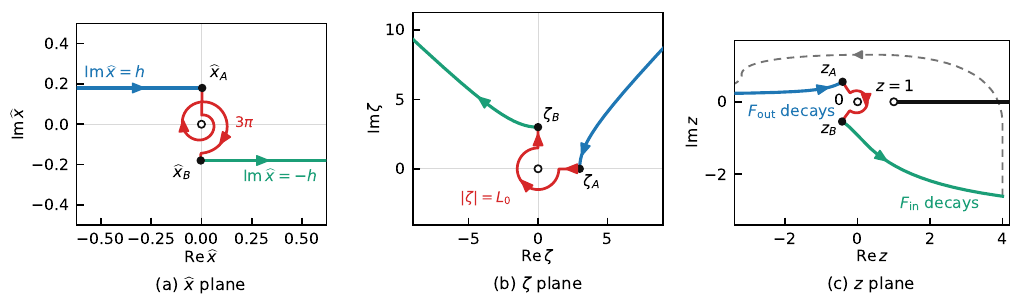}
\caption{The three parts of the horizon loop, drawn with exact maps for $\Omega=0.5-25\ii$, $L=3$, and $L_0=1.5$.
(a) In the $\widehat x$ plane, the outer contours are the lines $\operatorname{Im}\widehat x=\pm h$.
Towards infinity, $F_{\rm out}$ decays along the upper contour and $F_{\rm in}$ along the lower contour.
The inner arc has the constant radius $L_0^2/(2|\Omega|)$ and winds by $3\pi$ around $\widehat x=0$.
It is drawn with a growing radius for visibility.
The real parts of $\widehat x_A$ and $\widehat x_B$ are of order $w^{-1}$ and are not visible.
(b) In the $\zeta$ plane, the inner path consists of two half-rays and a clockwise arc of angle $3\pi/2$ on $|\zeta|=L_0$, and the outer contours are hyperbola branches.
(c) In the $z$ plane, the three parts and the closure at large $|z|$ (dashed, schematic) encircle the horizon $z=1$ counterclockwise.}
\label{supp:figpaths}
\end{figure}

The loop $\mathcal M_1$ is deformed into an upper contour, an inner path, and a lower contour, shown in Fig.~\ref{supp:figpaths}.
We call the upper and lower contours the outer contours.
In the $\widehat x$ plane, they are the lines $\operatorname{Im}\widehat x=h$ and $\operatorname{Im}\widehat x=-h$.
The upper contour runs from negative real infinity to the point $z_A$ with $\zeta_A=L$, so that $h=\operatorname{Im}[L^2/(2\Omega)]\simeq L^2/(2w)$.
The lower contour runs from the point $z_B$ with $\zeta_B=Le^{-3\pi\ii/2}$ on the lifted sheet to positive real infinity.
In the $\zeta$ plane, the inner path runs along the half-ray from $\zeta_A$ to the circle $|\zeta|=L_0$, along a clockwise arc of angle $3\pi/2$ on this circle, and along the half-ray to $\zeta_B$.
The outer contours keep a fixed distance from $z=0$ and $z=1$.
The inner path stays in the $z$ plane cut along the negative real axis.
From the positive real axis, where the waves \eqref{supp:boundary} are normalized, $F_{\rm out}$ is continued to the upper contour along an arc at large $|z|$ in the upper half plane.
This arc corresponds to the continuation of $F_{\rm out}$ through the right side of the frequency plane, since its decaying direction is $\arg z=\pi/2-\arg\Omega$.
With this arc the path is closed, and $\log(z-1)$ increases by $2\pi\ii$ along it.
The path therefore represents the counterclockwise loop about the outer horizon, as Fig.~\ref{supp:figpaths}(c) shows.
The coordinate $\widehat x$ is the one used in Ref.~\cite{MusiriSiopsis}, and the loop has the same topology as in the monodromy calculation of Ref.~\cite{MotlNeitzke}.
Our outer contours follow lines of monotone decay rather than lines on which the waves oscillate.

On the outer contours we compare $F_{\rm out}$ and $F_{\rm in}$ with the reference waves
\begin{equation}
J_\pm(z)=\left(\frac{\rho\Omega}{\overline k(z)}\right)^{1/2}
 e^{\pm\ii\overline S(z)},\qquad
\overline S(z)=\rho\Omega z+\Omega\log(z-1)+\beta\log z,\qquad
\overline k=f\,\frac{\dd\overline S}{\dd z}.
\label{supp:Jostreference}
\end{equation}
They retain the exact wave number and Coulomb power at infinity, and $\overline k=\Omega+O(w^{-1/2})$ on the outer contours.
Substitution into the wave equation gives $J_{\pm,\widehat x\widehat x}+(\Omega^2-V)J_\pm=\mathscr R J_\pm$, with
\begin{equation}
\mathscr R=\Omega^2-V-\overline k^2+\frac34\left(\frac{\overline k_{\widehat x}}{\overline k}\right)^2-\frac12\frac{\overline k_{\widehat x\widehat x}}{\overline k}.
\label{supp:JostResidual}
\end{equation}
The reference waves thus solve the wave equation with $V$ replaced by $V+\mathscr R$, and $\mathscr R$ plays the role of $g_\beta$ on the outer contours.
In $\Omega^2-V-\overline k^2$, the terms of order one and of order $1/z$ cancel exactly, and the remainder is $O(\sqrt w\,|z|^{-2})$.
The residual $\mathscr R$, divided by $\overline k$, therefore has an integral of order $w^{-1/2}$ along the outer contours.
Towards infinity, $e^{\ii\overline S}$ decays at the rate $w$ along the upper contour, and $e^{-\ii\overline S}$ along the lower contour.
On each outer contour, $|\overline k|$ is of order $w$, and the exponential factor of the Volterra kernel is bounded when we integrate in the decaying direction.
A Volterra equation along each outer contour then gives
\begin{equation}
F_{\rm out}=J_+\left[1+O(w^{-1/2})\right],\qquad
F_{\rm in}=J_-\left[1+O(w^{-1/2})\right],
\label{supp:outerWaveBounds}
\end{equation}
on the upper and lower contours, with the same relative errors for the derivatives.
These solutions are the canonical waves of \eqrefp{supp:boundary}, since a decaying solution with unit amplitude is unique.

At the two matching points, the definitions \eqrefp{supp:exactphase} and \eqrefp{supp:Jostreference} give
\begin{equation}
\overline S=\frac{\zeta^2}{2}\pm\ii\pi\Omega
+\beta\log\zeta-\beta\log(-\ii\sqrt\Omega)+O(w^{-1/2}),
\label{supp:phaseEndpoints}
\end{equation}
with the upper sign at $z_A$ and the lower sign at $z_B$.
The term $\pm\ii\pi\Omega$ comes from $\log(z-1)-\log(1-z)$, and the terms with $\beta$ come from $\log z$.
The same branch of $\log(-\ii\sqrt\Omega)$ appears in both cases.
The auxiliary equation \eqrefp{supp:innerBeta} has the unit-amplitude waves
\begin{equation}
y_{A,\beta}\sim\zeta^{-1/2+\ii\beta}e^{\ii\zeta^2/2},\qquad
y^-_{B,\beta}\sim\zeta^{-1/2-\ii\beta}e^{-\ii\zeta^2/2},
\label{supp:betaWaves}
\end{equation}
on the half-rays $\arg\zeta=0$ and $\arg\zeta=-3\pi/2$, respectively, up to relative terms of order $|\zeta|^{-1}$.
We denote by $\mathscr U_A$ and $\mathscr U_B$ the functions $\mathscr U$ obtained from $F_{\rm out}$ and $F_{\rm in}$.
At the matching points, \eqrefp{supp:outerWaveBounds} and \eqrefp{supp:phaseEndpoints} give, for values and derivatives,
\begin{equation}
\mathscr U_A=c_A\,y_{A,\beta}\left[1+O(w^{-1/2})\right],\quad
\mathscr U_B=c_B\,y^-_{B,\beta}\left[1+O(w^{-1/2})\right],\quad
c_A=(-\ii\sqrt\Omega)^{-\ii\beta},\quad
c_B=(-\ii\sqrt\Omega)^{\ii\beta}.
\label{supp:endpointMatch}
\end{equation}
Each prefactor is $1+O(w^{-1/2}\log w)$, but $c_Ac_B=1$ exactly.
We therefore keep the prefactors until the Wronskian is taken.

We transport these relations along the inner path $\Gamma$ to a common point on the arc.
Variation of constants turns the deviation $g_\beta$ into a Volterra integral equation, analogous to the Born series in scattering theory.
The solutions of \eqrefp{supp:innerBeta} and their derivatives are of order $|\zeta|^{-1/2}$ and $|\zeta|^{1/2}$ on the half-rays for $L_0\le|\zeta|\le L$.
In the integral equation, $g_\beta$ therefore appears divided by $1+|\zeta|$, the analogue of the wave number in the Born series.
Gronwall's inequality bounds the whole series, and the transport changes the relations at most by a constant times
\begin{equation}
\int_\Gamma\frac{|g_\beta(\zeta)|}{1+|\zeta|}\,|\dd\zeta|
\le C\left(\frac{L}{w}+w^{-1/2}\right)\le Cw^{-1/2},
\label{supp:innerIntegral}
\end{equation}
where $|\dd\zeta|$ denotes the arc length along $\Gamma$.
The bound follows from \eqrefp{supp:gbound}, and the arc contributes $O(w^{-1/2})$.

We denote by $T_{s,\beta}(\kappa_{\rm eff})=W_\zeta[y_{A,\beta}^{\rm cont},y^-_{B,\beta}]/(2\ii)$ the connection coefficient of the auxiliary equation, where the superscript denotes continuation along the arc.
At $\beta=0$, it is the coefficient $T_s$ of the Letter.
At the common point, the transported relations read $\mathscr U_A^{\rm cont}=c_Ay_{A,\beta}^{\rm cont}+O(w^{-1/2})$ and $\mathscr U_B=c_By^-_{B,\beta}+O(w^{-1/2})$, with the same errors for the derivatives.
Since $c_Ac_B=1$, we obtain
\begin{equation}
W_\zeta[\mathscr U_A^{\rm cont},\mathscr U_B]
=2\ii\,T_{s,\beta}(\kappa_{\rm eff})+O(w^{-1/2}).
\label{supp:Tbeta}
\end{equation}
Equation~\eqref{supp:innerBeta} maps to the sextic oscillator of Sec.~\ref{supp:B} with $\alpha=-4\ii\beta$ and $l=2|s|-1/2$, whose connection coefficients are analytic in $\alpha$.
Hence $T_{s,\beta}=T_s+O(w^{-1/2})$ on compact sets of $\kappa_{\rm eff}$.
Using \eqrefp{supp:exactLiouville} in \eqrefp{supp:factorization}, together with $\rho^{-1}=1+O(w^{-3/2})$, we obtain
\begin{equation}
\mathcal D(\Omega)
=e^{4\pi\Omega}+T_s(\kappa_{\rm eff})+O(w^{-1/2}).
\label{supp:matching}
\end{equation}
This is the matching estimate used in the Letter.

\subsection{Local frequencies}
We restrict $\xi$ to a compact set of positive values on which $T_s(\kappa_0)\ne0$ and the leading roots $-\ii w_N+\F_{0,s}(\xi)$ remain inside a fixed positive-real-part strip.
By Eq.~\eqref{supp:signRe}, the last condition holds on compact subsets of the intervals where $\operatorname{Re}\F_{0,s}>0$.
The first interval is $0<\xi<2\sqrt{\lambda_0}\simeq1.921$ for $s=0$ and $0<\xi<2\cdot6^{1/4}\simeq3.130$ for $s=-2$.
We hold $\chi_N=\xi w_N^{-3/4}$ fixed and vary $\Omega$ in a disk of fixed radius $r_0<1/2$ about $-\ii w_N+\F_{0,s}(\xi)$, inside the strip.
We use a normal prolate angular branch that is holomorphic in this disk and satisfies \eqrefp{supp:prolate} uniformly there.
As in \eqrefp{supp:kappalimit}, we then have $\kappa_{\rm eff}(\Omega)=\widehat\kappa_N+O(\eps^2)$ uniformly in the disk, where $\widehat\kappa_N=\kappa_0-\eps(2\bar L+1+2\ii m)\,\partial_\xi\kappa_0$ is independent of $\Omega$.
We compare $\mathcal D$ and the model function $\mathcal D_N^{(\kappa)}(\Omega)=e^{4\pi\Omega}+T_s(\kappa_{\rm eff}(\Omega))$ with
\begin{equation}
\widehat{\mathcal D}_N(\Omega)=e^{4\pi\Omega}+T_s(\widehat\kappa_N)
=T_s(\widehat\kappa_N)\left[1-e^{4\pi(\Omega-\widehat\Omega_N)}\right],
\qquad
\widehat\Omega_N=-\ii w_N+\frac{1}{4\pi}\log T_s(\widehat\kappa_N).
\label{supp:GN}
\end{equation}
The second expression follows from $e^{-4\pi\ii w_N}=-1$, and the logarithm is on the continuous branch of $\F_{0,s}$.
For large $N$, $\widehat{\mathcal D}_N$ has the single simple zero $\widehat\Omega_N$ in the disk, and $|\widehat{\mathcal D}_N|$ is bounded below by a constant times $|\Omega-\widehat\Omega_N|$.
The functions $\mathcal D_N^{(\kappa)}$ and $\mathcal D$ are holomorphic there.
They differ from $\widehat{\mathcal D}_N$ by $O(\eps^2)$, by the analyticity of $T_s$ and by \eqrefp{supp:matching}.
By Rouch\'e's theorem, each of them has exactly one zero in the disk, and this zero lies within $O(\eps^2)$ of $\widehat\Omega_N$.
We denote these zeros by $\Omega_N^{(\kappa)}$ and $\Omega_N$.
The latter is a standard Teukolsky QNM frequency, and
\begin{equation}
\Omega_N-\Omega_N^{(\kappa)}
=O\big(w_N^{-1/2}\big).
\label{supp:rootBound}
\end{equation}
This is a statement about selected local roots.
Their relation to the conventional overtone number $n$ is checked numerically in Sec.~\ref{supp:F}.
The expansion of the logarithm in \eqrefp{supp:GN} then yields
\begin{equation}
\begin{aligned}
\Omega_N+\ii w_N
&=\F_{0,s}(\xi)+\eps\F_{1,s,m,\bar L}(\xi)
+O(\eps^2),\\
\F_{0,s}(\xi)&=\frac1{4\pi}\log T_s(\kappa_0),\\
\F_{1,s,m,\bar L}(\xi)&=-(2\bar L+1+2\ii m)\,\partial_\xi\F_{0,s}(\xi).
\end{aligned}
\label{supp:frequencyExpansion}
\end{equation}
We do not claim uniformity for $\xi\to\infty$ or physical continuation through the negative imaginary frequency axis.

\section{Spectral determinants and asymptotic expansions}\label{supp:E}
\subsection{Endpoint conditions and the determinant}
The matching of Sec.~\ref{supp:D} involves the coefficient $T_s$ of the dual equation, and we compute it with the dual oscillator.
By Eq.~\eqref{supp:Tidentity}, the standard oscillator gives the same coefficient.
We write $\nu=|s|$, $l_{\rm d}=2\nu-1/2$, and $X=e^{\ii\pi/4}\kappa$, so $E=2\sqrt2X$.
On $x>0$, we select the regular endpoint solution $\varphi_{\rm reg}\sim x^{l_{\rm d}+1}$ and the decaying solution $Y^{(1)}\sim x^{-3/2}e^{-x^4/4}$.
The superscript on $Y^{(1)}$ denotes unit leading amplitude, while subscripts label sectors.
It differs from \eqrefp{supp:Y} at $\alpha=0$ by $\sqrt{2\ii}$.
For $s=0$, the solutions at the origin behave as $x^{1/2}$ and $x^{1/2}\log x$, and we take the former.
The spectral determinant and its normalized version are
\begin{equation}
D_s(E)=W_x[Y^{(1)},\varphi_{\rm reg}],\qquad
Q_s(X)=\frac{D_s(2\sqrt2X)}{D_s(0)}.
\label{supp:Ddefinition}
\end{equation}
The endpoint solutions become dependent at the eigenvalues $E_k=2\sqrt2\lambda_k$.
The function $x^{-1/2}\varphi$ obeys the radial equation of the two-dimensional Schr\"odinger operator $-\nabla^2+x^6$, with radius $x$ and angular momentum $2\nu$.
The solution $\varphi_{\rm reg}$ corresponds to its regular solution.
Since this operator is positive, the eigenvalues are positive and $D_s(0)\ne0$.

WKB counting gives $E_k\sim C k^{3/2}$ with a positive constant $C$.
This ensures convergence of $\prod_k(1-X/\lambda_k)$.
The determinant has order $2/3<1$~\cite{DoreyTateo}, so Hadamard factorization and $Q_s(0)=1$ give
\begin{equation}
Q_s(X)=\prod_{k=0}^{\infty}\left(1-\frac{X}{\lambda_k}\right).
\label{supp:Qproduct}
\end{equation}
The product converges at finite $X$, while the large $X$ series below is asymptotic.

Taking the Wronskian of \eqrefp{supp:basicC} with the origin solution gives
\begin{equation}
C_{\rm d}(E)Q_s(X)=e^{-\ii\pi\nu/2}Q_s(-\ii X)+e^{\ii\pi\nu/2}Q_s(\ii X).
\label{supp:TQdual}
\end{equation}
The twist follows from the origin power under \eqrefp{supp:rotated}.
The standard deformed problem has the two-determinant relation of Ref.~\cite{DoreyDunningTateo}, Eqs.~(2.10)--(2.11).
We use the dual relation above, which avoids normalization by the zero standard determinant at $s=-1$, $E=0$.

Combining \eqrefp{supp:TQdual} with \eqrefp{supp:dualFusion}, we define
\begin{equation}
\mathcal A_s(X)=\frac{Q_s(\ii X)}{Q_s(-\ii X)},\qquad
\mathcal B_s(X)=\frac{Q_s(X)}{Q_s(-X)}
\label{supp:AB}
\end{equation}
and obtain
\begin{equation}
T_s(\kappa)+1=[1+e^{\ii\pi\nu}\mathcal B_s(X)]
[1+e^{-\ii\pi\nu}\mathcal A_s(X)].
\label{supp:Tproduct}
\end{equation}
The algebra is first performed where the denominators are nonzero.
Elsewhere removable singularities are understood through the entire Wronskian expression \eqref{supp:Tbridge}.
On positive real $X$, the denominators in \eqrefp{supp:AB} do not vanish.
These identities follow from the differential equation alone.
They hold for the gravitational parameter $l_{\rm d}=7/2$, which lies outside the basic range of the nonlinear integral equation of Ref.~\cite{DoreyTateo}.

Reference~\cite{JiaRangamani} derives the Stokes multipliers of a scaled equation near the singularity of Schwarzschild black holes in anti-de Sitter space, and writes them as ratios of determinants.
The equation is stated there for odd spacetime dimensions, five and higher.
Its formal continuation to four dimensions becomes Eq.~\eqref{supp:innerD} with $s=0$, where $\kappa$ is the square of their scaled momentum.
For this, we write their radial variable as $v^2$ and divide their wave function by $(2v)^{1/2}$.
The determinant then becomes $Q_0$, and the product of two adjacent multipliers becomes $T_0+1$ as given by Eq.~\eqref{supp:Tproduct}.
Their determinant formula has the form of Eq.~\eqref{supp:TQdual}, with quartic and cubic potentials in five and seven dimensions.
The Kerr realization of this local problem and its matching to the QNM boundary conditions are derived in Secs.~\ref{supp:C} and \ref{supp:D}.

For $s=0,-2$ and positive real $X$, the eigenvalues are positive, so that $|\mathcal A_s|=1$ and $\mathcal B_s$ is real with $|\mathcal B_s|<1$.
Equation~\eqref{supp:Tproduct} then gives
\begin{equation}
|T_s|^2-1=2\mathcal B_s(1+\mathcal B_s)(1+\operatorname{Re}\mathcal A_s).
\label{supp:signRe}
\end{equation}
For $\mathcal A_s\ne-1$, the real part of $\F_{0,s}$ therefore has the sign of $\mathcal B_s$, which is the sign of $Q_s(X)$.
At $X=\lambda_k$, Eq.~\eqref{supp:TQdual} gives $\mathcal A_s=-1$, so that $T_s=-1$ and $\F_{0,s}=-\ii/4$ modulo $\ii/2$.
The real part changes sign at these points.
The first one is $\xi=2\sqrt{\lambda_0}\simeq1.921$ for $s=0$ and $\xi=2\sqrt{\lambda_0}\simeq5.565$ for $s=-2$.
For $s=-2$, we also have $T_s=-1$ at the quasi-exactly solvable energy $E=4\sqrt3$, that is, at $\xi=2\cdot6^{1/4}\simeq3.130$.
Indeed, $x^{-7/2}(1+Ex^2/12)e^{x^4/4}$ solves $H_{\rm d}\varphi=E\varphi$ for $E^2=48$ and decays in both $\mathcal S_1$ and $\mathcal S_{-1}$.
It is proportional to $Y_1$ and to $Y_{-1}$, so that $C_{\rm d}(\pm4\sqrt3)=0$ by Eq.~\eqref{supp:basicC}.
Equation~\eqref{supp:dualFusion} then gives $T_s=-1$.
There $\mathcal A_s=-1$ with $\mathcal B_s>0$, and the real part touches zero without changing sign.
At all these points the leading frequency lies on the negative imaginary axis, outside the domain \eqref{supp:P1domain}.

\subsection{The Schwarzschild endpoint}
At $X=0$, Eq.~\eqref{supp:Tproduct} gives $T_s(0)=1+2\cos(\pi s)$.
It equals $3$ for $s=0,-2$ and $-1$ for $s=-1$.
As an independent normalization check, the dual equation at $\kappa=0$ reduces to a Bessel equation of index $|s|/2$ in the variable $v^2/2$.
Clockwise continuation of $v$ by $3\pi/2$ gives $T_s(0)=\sin(3\pi|s|/2)/\sin(\pi|s|/2)$, continued at integer indices~\cite{MotlNeitzke}.

For $|X|<\lambda_0$, the product gives the convergent expansion
\begin{equation}
\log Q_s(X)=-\sum_{r\ge1}\frac{Z_{r,s}}rX^r,
\qquad Z_{r,s}=\sum_{k\ge0}\lambda_k^{-r}.
\label{supp:zetas}
\end{equation}
The first sum follows from the inverse spectral sum rule of Watkins for the sextic potential~\cite{Watkins},
\begin{equation}
Z_{1,s}=\frac{\sqrt\pi\,\Gamma(1/4)\Gamma(1/4+|s|/2)}
{8\Gamma(3/4)\Gamma(3/4+|s|/2)}.
\label{supp:Z1}
\end{equation}
Its normalization is $Z_{1,s}=2\sqrt2\sum_k E_k^{-1}$ because $E_k=2\sqrt2\lambda_k$.
For the determinant computation we also use the equivalent weighted problem on $v=e^{\ii\pi/4}t$, $t>0$,
\begin{equation}
\left[-\partial_t^2+t^2+\frac{\nu^2-1/4}{t^2}\right]u=Xt^{-1}u.
\label{supp:weighted}
\end{equation}
It has the regular behavior $t^{\nu+1/2}$ at zero and a decaying solution at infinity.
For $s=0,-2$, expansion of Eq.~\eqref{supp:Tproduct} yields
\begin{equation}
T_s(\kappa_0)=3-4(1+\ii)Z_{1,s}X+O(X^2),\qquad
\F_{0,s}(\xi)=\frac{\log3}{4\pi}-\frac{(1+\ii)Z_{1,s}}{12\pi}\xi^2+O(\xi^4),
\quad X=\frac{\xi^2}{4}.
\label{supp:smallxi}
\end{equation}
The logarithm is continued from $\log3$ along a zero-free path.

\subsection{The large parameter expansion}
For positive $X$, $|\mathcal A_s|=1$ and $\mathcal B_s$ is real.
The latter is exponentially small for large $X$.
Indeed, every factor of
$|\mathcal B_s|=\prod_k|\lambda_k-X|/(\lambda_k+X)$ is at most one.
For $2X\leq\lambda_k\leq4X$, it is at most $3/5$.
The number of such eigenvalues grows as a positive constant times $X^{2/3}$.
Thus $|\mathcal B_s(X)|\leq e^{-c_*X^{2/3}}$ for sufficiently large $X$ and some $c_*>0$.
For the even spin weights in the Letter,
\begin{equation}
T_s=\mathcal A_s+\mathcal B_s(1+\mathcal A_s),\qquad
\log T_s=\log\mathcal A_s+O(e^{-c_*X^{2/3}}),
\label{supp:largeA}
\end{equation}
up to the consistently transported logarithmic constant.

Rescaling $x=E^{1/6}t$ in the pure sextic equation gives
\begin{equation}
\left[-E^{-4/3}\partial_t^2+t^6+E^{-4/3}\frac{l_{\rm d}(l_{\rm d}+1)}{t^2}\right]\varphi=\varphi.
\label{supp:WKBscale}
\end{equation}
The effective WKB parameter is $E^{-2/3}$.
The local action series therefore contains powers with exponents
$\alpha_r=\frac23(1-2r)$, $r=0,1,\ldots$.
For complex $X$, a local part of the determinant asymptotics can be written as
\begin{equation}
\log Q_s(X)\big|_{\rm local}\sim
\sum_{r\geq0}b_r(s)(-X)^{\alpha_r}-\nu\log(-X)+b_{\rm const}.
\label{supp:logQlarge}
\end{equation}
The leading complex WKB asymptotics is valid in $|\arg(-X)|<\pi$~\cite{DoreyTateo}.
We apply it at $\pm\ii X$ with $X>0$, which lie inside that sector.
For each algebraic term,
\begin{equation}
(-\ii X)^{\alpha_r}-(\ii X)^{\alpha_r}
=-2\ii\sin\left(\frac{\pi\alpha_r}{2}\right)X^{\alpha_r}.
\label{supp:rotationfactor}
\end{equation}
By Eqs.~\eqref{supp:largeA} and \eqref{supp:logQlarge}, the algebraic part of the expansion of $\F_{0,s}$ is
\begin{equation}
\F_{0,s}(\xi)\sim\frac{\ii\nu}{4}-\frac{\ii}{2}\sum_{r\ge0}a_r(s)X^{\alpha_r},\qquad
a_r(s)=\frac{b_r(s)}{\pi}\sin\frac{\pi\alpha_r}{2},\qquad X=\frac{\xi^2}{4}.
\label{supp:arDef}
\end{equation}
The constant comes from $-\nu[\log(-\ii X)-\log(\ii X)]=\ii\pi\nu$.

\subsection{The dictionary to the integrable model}
The integrable model is the integrable structure of conformal field theory (CFT) constructed by Bazhanov, Lukyanov, and Zamolodchikov (BLZ)~\cite{BLZI,BLZII}.
For the sextic oscillator, the dictionary of Ref.~\cite{DoreyTateo} between the differential equation and the integrable model gives
\begin{equation}
\beta_{\rm BLZ}^2=\frac14,\qquad p_{\rm CFT}=\frac\nu4,\qquad
c_{\rm CFT}=1-6(\beta_{\rm BLZ}-\beta_{\rm BLZ}^{-1})^2=-\frac{25}{2},\qquad
\Delta_{\rm CFT}=\frac{p_{\rm CFT}^2}{\beta_{\rm BLZ}^2}+\frac{c_{\rm CFT}-1}{24}
=\frac{s^2}{4}-\frac9{16}.
\label{supp:BLZdictionary}
\end{equation}
Here $c_{\rm CFT}$ is the central charge, distinct from $c=a\omega$.
The vacuum eigenvalue refers to the highest weight state of weight $\Delta_{\rm CFT}$, which differs from the identity state for $s=0,-2$.
The normalized determinant is identified with the vacuum eigenvalue $A_+$ of the BLZ Baxter operator~\cite{BLZspec},
\begin{equation}
Q_s(X)=A_+(\lambda_{\rm BLZ},p_{\rm CFT}),\qquad
\lambda_{\rm BLZ}^2=\frac{X}{8\Gamma(3/4)^2}.
\label{supp:BLZscale}
\end{equation}
This follows from $\lambda_{\rm BLZ}=8^{-3/4}E^{1/2}/\Gamma(3/4)$ in Ref.~\cite{DoreyTateo}, Eq.~(2.26), and $E=2\sqrt2X$.
The scalar endpoint uses its log-free limit.

The BLZ expansion organizes the local coefficients through commuting local integrals of motion $I_{2r-1}$~\cite{BLZI,BLZII}.
We denote their highest weight eigenvalues by $I_{2r-1}^{\rm vac}(\Delta_{\rm CFT})$.
The coefficients of Eq.~\eqref{supp:arDef} with $r\ge1$ are
\begin{equation}
a_r(s)=\mathcal K_r I_{2r-1}^{\rm vac}(\Delta_{\rm CFT}),\qquad
\mathcal K_r=-\frac4{3\sqrt\pi}\sin\frac{\pi(1-2r)}3
\frac{4^{r-1}\Gamma((4r-2)/3)}{r!\,\Gamma((2r+5)/6)},\qquad r\geq1.
\label{supp:Kr}
\end{equation}
The factor $\sin[\pi(1-2r)/3]$ is the sine factor of Eq.~\eqref{supp:arDef}.
The normalization can be traced from Eq.~(4.16) of Ref.~\cite{BLZII}.
At $\beta_{\rm BLZ}^2=1/4$, its transfer-matrix coefficient is
\begin{equation}
C_r^{\rm BLZ}=\frac{4\sqrt\pi}{3r!}\,4^{-r}
\frac{\Gamma((4r-2)/3)}{\Gamma((2r+5)/6)}
\Gamma(3/4)^{-\frac43(2r-1)}.
\label{supp:BLZCn}
\end{equation}
Writing the local term of $\log A_+$ as
$c_r^{\rm BLZ}(-\lambda_{\rm BLZ}^2)^{\alpha_r}I_{2r-1}^{\rm vac}$,
the dominant term of the asymptotic T--Q relation gives
\begin{equation}
c_r^{\rm BLZ}=-\frac{C_r^{\rm BLZ}}
{2(-1)^r\sin(\pi\alpha_r/4)},\qquad
\mathcal K_r=\frac{c_r^{\rm BLZ}}{\pi}
[8\Gamma(3/4)^2]^{-\alpha_r}\sin\frac{\pi\alpha_r}{2}.
\label{supp:BLZconversion}
\end{equation}
The denominator sine is nonzero for integer $r\geq1$, and substituting $\alpha_r=2(1-2r)/3$ reduces this expression to \eqrefp{supp:Kr}.
The expansion rests on the asymptotic structure of Ref.~\cite{BLZII}, whose Eq.~(4.19) relies on the analytic conjectures stated there.

The first eigenvalue is
$I_1^{\rm vac}=\Delta_{\rm CFT}-c_{\rm CFT}/24=(s^2-1/6)/4$.
With $\mathcal K_1=4B_1/(2^{7/3}\pi)$, it reproduces the first inverse-power term,
\begin{equation}
\F_{0,s}(\xi)\sim-\frac{3\ii B_0}{32\pi}\xi^{4/3}
+\frac{\ii\nu}{4}
+\frac{\ii(1/6-s^2)B_1}{4\pi}\xi^{-4/3}+\cdots,
\quad
B_0=\frac{2\,4^{-1/6}}{\sqrt3}\mathrm B\left(\frac13,\frac12\right),\quad
B_1=\frac{2\,4^{1/6}}{\sqrt3}\mathrm B\left(\frac23,\frac12\right).
\label{supp:firstlarge}
\end{equation}
Here $\mathrm B(a,b)=\Gamma(a)\Gamma(b)/\Gamma(a+b)$ is the Euler beta function.
The term with $r=0$ is a classical action, with $a_0=3\cdot4^{2/3}B_0/(16\pi)$.
We retain the gravitational constant $\ii/2$ rather than shifting the lattice afterward.

\begin{figure}[t]
\centering
\includegraphics[width=0.92\textwidth]{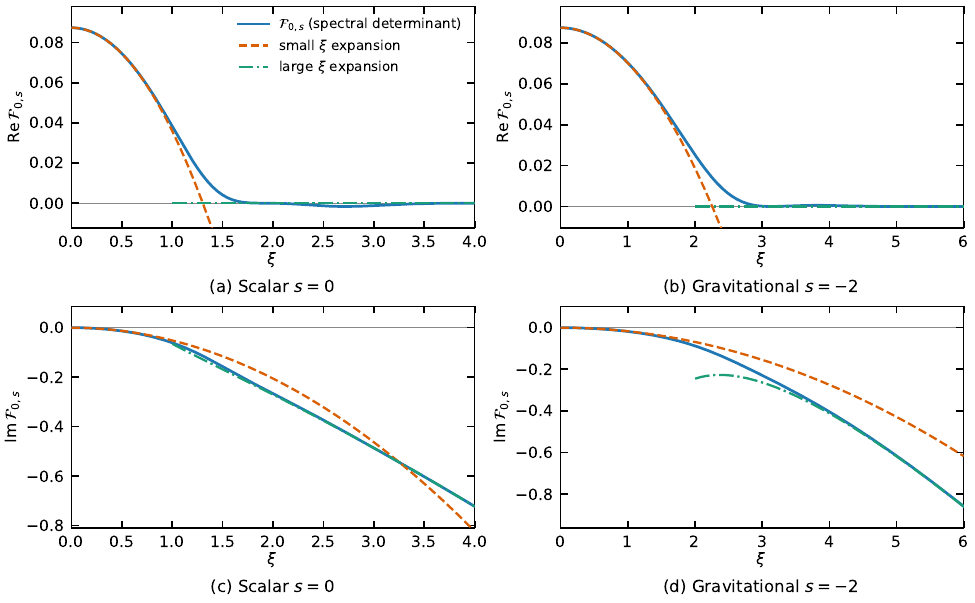}
\caption{The crossover function $\F_{0,s}(\xi)$ and its two limits.
Solid curves show $\F_{0,s}$ computed from the spectral determinant, as in Fig.~\ref{fig:crossover}.
Dashed curves show the small $\xi$ expansion, Eq.~\eqref{eq:small}.
Dash-dotted curves show the large $\xi$ expansion, Eq.~\eqref{eq:firstlarge}, which has no real part.
Our derivation of Eq.~\eqref{eq:expansion} covers the values of $\xi$ with $\operatorname{Re}\F_{0,s}>0$, in particular $\xi<1.92$ for $s=0$.
Elsewhere, the figure shows the function $\F_{0,s}$ itself.}
\label{supp:figasym}
\end{figure}

Figure~\ref{supp:figasym} compares Eqs.~\eqref{eq:small} and \eqref{eq:firstlarge} with $\F_{0,s}$ computed from the spectral determinant.
The error of Eq.~\eqref{eq:small} is below $10^{-2}$ for $\xi<1.0$ when $s=0$ and for $\xi<1.7$ when $s=-2$.
The error of the imaginary part of Eq.~\eqref{eq:firstlarge} is below $10^{-2}$ for $\xi>1.6$ and $\xi>3.9$, respectively.
Following the phase of $T_s$ continuously from $\xi=0$, we find that the imaginary part of $\F_{0,s}$ differs from that of Eq.~\eqref{eq:firstlarge} by less than $10^{-2}$ for $4\le\xi\le35$ and both $s$.
This comparison shows that the constant $\ii\nu/4$ belongs to the branch continued from $\log3$, without a shift by a multiple of $\ii/2$.

We compare Eq.~\eqref{supp:arDef} with the expansion at fixed rotation of Ref.~\cite{Paper1}, $\Omega_N=\sum_{j\ge-1}d_j(\chi)(N+1/2)^{-j}$.
The integer $N$ of Ref.~\cite{Paper1} is defined by the quantization condition at fixed rotation.
It agrees with ours for $s=0$, and it is smaller than ours by one for $s=-2$, where the constant $\ii\nu/4=\ii/2$ is absorbed in this shift.
The shift does not change the terms proportional to $\chi^{-4j/3}$.
With $j=2r-1$, we have $X^{\alpha_r}=2^{7j/3}\chi^{-4j/3}(N+1/2)^{-j}$.
The term $r$ of Eq.~\eqref{supp:arDef} thus contributes $-\ii\,2^{7j/3-1}a_r(s)\chi^{-4j/3}$ to $d_j$.
These contributions agree with the terms of $d_j$ proportional to $\chi^{-4j/3}$ obtained in Ref.~\cite{Paper1}.
The term $\eps\F_{1,s,m,\bar L}$ of Eq.~\eqref{supp:frequencyExpansion} gives the terms of $d_0$ and $d_2$ proportional to $\chi^{1/3}$ and $\chi^{-7/3}$.
For $(s,m,\bar L)=(-2,-2,0)$, the latter is $(-24.49361921-6.12340480\,\ii)\chi^{-7/3}$, in agreement with Eq.~(3.51) of Ref.~\cite{Paper1}.
The sine factor vanishes for $r=2,5,8,\ldots$.
Hence $d_j$ with $j\equiv3\pmod6$ has no term proportional to $\chi^{-4j/3}$, and in particular $d_3$ has no term proportional to $\chi^{-4}$.
Its leading term is proportional to $\chi^{-10/3}$~\cite{Paper1}.
In the scaling variables, this term is of order $\eps^2$, since $\chi^{-10/3}(N+1/2)^{-3}=\eps^2\xi^{-10/3}/8$.
The term $\eps\F_{1,s,m,\bar L}$ does not contribute to it, since $\F_{1,s,m,\bar L}$ is proportional to $\partial_\xi\F_{0,s}$ and contributes only to $d_j$ with even $j$.

The expansion of Ref.~\cite{BLZII} also has a second, nonlocal series.
Its powers are $X^{-4n}$ with $n\geq1$, and they cancel in $\log Q_s(\ii X)-\log Q_s(-\ii X)$.

\subsection{Argyres--Douglas limit of the \texorpdfstring{$N_f=3$}{Nf=3} curve}
\label{supp:AD}
The Kerr radial equation is a quantum Seiberg--Witten curve of $SU(2)$ theory with three fundamental hypermultiplets~\cite{AGH_SW},
\begin{equation}
\hbar^2\Psi''+\frac{\mathcal P_4(z)}{z^2(z-1)^2}\Psi=0.
\label{supp:AD_SW}
\end{equation}
Here $\mathcal P_4$ is a quartic polynomial, whose coefficients are given in Eqs.~(2.27)--(2.29) of Ref.~\cite{AGH_SW}.
For the standard radial equation, their dictionary in units $2M=1$ is
\begin{equation}
\begin{aligned}
\Lambda_3&=-8\ii\rho\Omega,\qquad \rho=\sqrt{1-\chi^2},\\
\mathfrak m_1&=s-\ii\Omega,\qquad
\mathfrak m_2=\frac{\ii(-\Omega+\chi m)}\rho,\qquad
\mathfrak m_3=-s-\ii\Omega,\\
\mathcal E&=-A(c)-s(s+1)+\left(2-\frac{\chi^2}{4}\right)\Omega^2-\frac14.
\end{aligned}
\label{supp:AD_dictionary}
\end{equation}
The gauge masses $\mathfrak m_i$ differ from the azimuthal number $m$, and $\mathcal E$ is the Coulomb parameter.
The dual assignment exchanges $\mathfrak m_2$ and $\mathfrak m_3$ and uses its own Heun coordinate.
At fixed positive $\xi$, both assignments give
\begin{equation}
\frac{\mathfrak m_i}{\Lambda_3}\longrightarrow\frac18,
\qquad \frac{\mathcal E}{\Lambda_3^2}\longrightarrow-\frac1{32}.
\label{supp:AD_parameters}
\end{equation}
These are the $(A_1,D_4)$ Argyres--Douglas ratios~\cite{AGH_SW,ItoOkubo_AD,ItoKoizumiOkubo_AD}.
At the exact classical locus, substitution into these coefficients yields
\begin{equation}
\mathcal P_4(z)\big|_{\rm cl}=-\frac{\Lambda_3^2}{64}\,z^4.
\label{supp:AD_degeneration}
\end{equation}
The four classical turning points coalesce at $z=0$.
Their quantum scaling is resolved by the two inner equations \eqref{supp:innerT} and \eqref{supp:innerD}, each in its own radial coordinate.
At $\hbar=1$, the inverse square coefficient $[1-(\mathfrak m_1-\mathfrak m_2)^2]/4$ tends to $(1-s^2)/4$ or $1/4-s^2$ in the standard or dual assignment.
The $1/4$ terms are quantum corrections rather than classical masses.
The quantum $D$-type curve depends on the quantization convention~\cite{ItoKoizumiOkubo_AD}.

In the local classical curve, the Seiberg--Witten differential is $\lambda_{\rm SW}=(v^2+\kappa_0/v)^{1/2}\,\dd v$.
We denote by $[\,\cdot\,]$ the scaling dimension, fixed by $[\lambda_{\rm SW}]=1$.
Thus $[v]=1/2$ and $[\kappa_0]=3/2$.
At nonzero $\xi$, $\kappa_0$ remains a deformation of the fixed point curve.

\section{Numerical methods and verification}\label{supp:F}\label{supp:data}
Parts of the numerical code were written with AI assistance, and all results were checked as described below.
We distinguish physical spectrum tests from checks of the boundary coefficients and of the limiting equations.
Only the first group uses full Kerr QNM roots.
The boundary tests probe the matching relation and its normalization without imposing a root condition.
The sequences of the second subsection were computed separately for Fig.~\ref{fig:crossover} of the Letter.

\subsection{Physical frequencies and model roots}
The standard Teukolsky radial continued fraction is solved together with the angular eigenvalue condition~\cite{Leaver,Nollert}.
The radial ansatz is
\begin{equation}
R_{\rm T}=e^{\ii\omega r}(r-r_-)^{-1-s+\ii\Omega+\ii\sigma_+}
(r-r_+)^{-s-\ii\sigma_+}\sum_{j\ge0}a_j
\left(\frac{r-r_+}{r-r_-}\right)^j.
\label{supp:Jaffe}
\end{equation}
The minimal solution of the resulting recurrence $\alpha_j^r a_{j+1}+\beta_j^r a_j+\gamma_j^r a_{j-1}=0$ gives
\begin{equation}
\beta_0^r-\frac{\alpha_0^r\gamma_1^r}{\displaystyle\beta_1^r-
\frac{\alpha_1^r\gamma_2^r}{\displaystyle\beta_2^r-\cdots}}=0.
\label{supp:CF}
\end{equation}
The coefficients follow Eqs.~(18)--(27) of Ref.~\cite{Leaver}, with $2M\omega=\Omega$ and $a/(2M)=\chi/2$.
We use inverted fractions and an asymptotic tail.
The inversion index is distinct from $n$ and $N$.
The angular fraction is checked against a spin weighted spherical harmonic matrix calculation.

Table~\ref{supp:tabRoots} contains all 24 physical frequencies and the differences from the separately solved model.
For each case, we repeat the root search 16 times with different settings.
We vary the arithmetic from 44 to 88 digits, the depths and tails of the continued fractions, the inversion index, the cutoff of the angular matrix, and the starting values.
The frequencies change by at most $6.8\times10^{-16}$, which we take as their numerical uncertainty.

We define the overtone number $n$ at $\chi=0$ by counting the Schwarzschild frequencies in order of increasing damping.
For the gravitational modes, the count includes the algebraically special frequency as $n=8$.
For each of the 24 cases, we continue the Schwarzschild frequencies with $n=N$, $N+1$, and $N-1$ in $\chi$ from $0$ to $\chi_N$.
The continuation from $n=N$ arrives at the frequency of Table~\ref{supp:tabRoots}, and the continuation back to $\chi=0$ returns to its starting point.
None of these paths crosses the negative imaginary frequency axis.

\begin{table}[htbp]
\centering\small
\caption{The 24 standard Teukolsky frequencies and differences from independent model roots. We display $\delta_N=\Omega_N+\ii w_N$ to avoid printing the large common damping term. All values are rounded to nine decimals, which is much coarser than their numerical uncertainty. For this set, $N$ equals the overtone number $n$.}\label{supp:tabRoots}
\begin{tabular}{rrrrrrr}
\hline\hline
$s$ & $\xi$ & $N$ & $\operatorname{Re}\delta_N$ & $\operatorname{Im}\delta_N$ & $\operatorname{Re}\Delta_{\kappa,N}$ & $\operatorname{Im}\Delta_{\kappa,N}$\\
\hline
$0$ & $0.5$ & $400$ & $0.083566213$ & $-0.001037903$ & $0.004866330$ & $0.007953277$ \\
$0$ & $0.5$ & $800$ & $0.082067856$ & $-0.003405512$ & $0.003435076$ & $0.005651345$ \\
$0$ & $0.5$ & $1600$ & $0.080968824$ & $-0.005139005$ & $0.002425797$ & $0.004010326$ \\
$0$ & $0.5$ & $3200$ & $0.080136684$ & $-0.006431891$ & $0.001713598$ & $0.002843417$ \\
$0$ & $1$ & $400$ & $0.056016970$ & $-0.023799383$ & $0.004505396$ & $0.017644117$ \\
$0$ & $1$ & $800$ & $0.053937796$ & $-0.029890475$ & $0.003177840$ & $0.012561593$ \\
$0$ & $1$ & $1600$ & $0.052088539$ & $-0.034765894$ & $0.002237862$ & $0.008932394$ \\
$0$ & $1$ & $3200$ & $0.050384651$ & $-0.038800226$ & $0.001573871$ & $0.006346896$ \\
$0$ & $1.5$ & $400$ & $0.015648975$ & $-0.081674400$ & $0.001897986$ & $0.032851661$ \\
$0$ & $1.5$ & $800$ & $0.013866982$ & $-0.095040869$ & $0.001355677$ & $0.023467344$ \\
$0$ & $1.5$ & $1600$ & $0.012216346$ & $-0.106038598$ & $0.000937349$ & $0.016706578$ \\
$0$ & $1.5$ & $3200$ & $0.010770694$ & $-0.115101968$ & $0.000636215$ & $0.011865349$ \\
$-2$ & $0.5$ & $400$ & $0.130904807$ & $-0.003148201$ & $-0.010897194$ & $-0.001185542$ \\
$-2$ & $0.5$ & $800$ & $0.122822160$ & $-0.006183242$ & $-0.006913848$ & $-0.000915952$ \\
$-2$ & $0.5$ & $1600$ & $0.116055126$ & $-0.007785844$ & $-0.004447329$ & $-0.000710852$ \\
$-2$ & $0.5$ & $3200$ & $0.110457504$ & $-0.008416030$ & $-0.002895414$ & $-0.000543256$ \\
$-2$ & $1$ & $400$ & $0.147318867$ & $-0.020509370$ & $-0.016686164$ & $0.009799081$ \\
$-2$ & $1$ & $800$ & $0.137708647$ & $-0.021943528$ & $-0.010143942$ & $0.006378529$ \\
$-2$ & $1$ & $1600$ & $0.128634579$ & $-0.021195313$ & $-0.006309361$ & $0.004209357$ \\
$-2$ & $1$ & $3200$ & $0.117197351$ & $-0.018472372$ & $-0.003998810$ & $0.002771936$ \\
$-2$ & $2$ & $400$ & $0.148644659$ & $-0.026890010$ & $-0.016836427$ & $0.048933782$ \\
$-2$ & $2$ & $800$ & $0.131652861$ & $-0.043483929$ & $-0.010672163$ & $0.031063290$ \\
$-2$ & $2$ & $1600$ & $0.116068130$ & $-0.056153565$ & $-0.006796554$ & $0.020694818$ \\
$-2$ & $2$ & $3200$ & $0.102141868$ & $-0.065660997$ & $-0.004290630$ & $0.014107679$ \\
\hline\hline
\end{tabular}
\end{table}

The model roots $\Omega_N^{(\kappa)}$ are the zeros of the function $\mathcal D_N^{(\kappa)}$ of Sec.~\ref{supp:D}, with the angular eigenvalue at finite $c$.
We solve for them without using the physical roots.
We start from the Schwarzschild root at $\chi=0$ with $\kappa_{\rm eff}$ replaced by zero, and continue first in a factor multiplying $\kappa_{\rm eff}$ and then in $\chi$.

We evaluate $T_s$ from its Taylor series in $\kappa$, with coefficients computed by Cauchy integrals of direct integrations of the inner equation on $|\kappa|=2$.
The values of $|\kappa_{\rm eff}|$ at the model roots are at most $1.33$.
Two settings of this calculation, with 42 and 60 digit arithmetic, change the model roots by at most $1.2\times10^{-29}$.

Fits of $\log|\Delta_{0,N}|$ and $\log|\Delta_{\kappa,N}|$ against $\log N$ for $N=400$ to $3200$ give slopes from $-0.28$ to $-0.23$ and from $-0.66$ to $-0.49$, respectively.
They agree with the bounds of order $\eps$ and $\eps^2$, as does the longer sequence in Sec.~\ref{supp:extrap}.
For gravity at $\xi=0.5$, $|c|$ is only about $0.94$ to $1.58$ at these points, so that the angular eigenvalue is still in its transition.

We compare separately solved roots rather than the value of $\mathcal D_N^{(\kappa)}$ at the physical frequency.
The two are related by $\Delta_{\kappa,N}\simeq\mathcal D_N^{(\kappa)}(\Omega_N)/\partial_\Omega\mathcal D_N^{(\kappa)}(\Omega_N)$, and the modulus of the derivative ranges from $14.39$ to $81.36$ for the 24 frequencies.

\subsection{Extrapolation in the overtone number}\label{supp:extrap}
The points in Fig.~\ref{fig:crossover} of the Letter come from sequences at larger $N$ than those of Table~\ref{supp:tabRoots}.
For each $\xi$, we solve Eq.~\eqref{supp:CF} together with the angular eigenvalue condition at 20 to 24 values of $N$, with $\chi_N=\xi w_N^{-3/4}$.
Table~\ref{supp:tabExtrap} lists the ranges of $N$.
The rotation $\chi_N$ ranges from $7\times10^{-5}$ to $0.014$.
For the gravitational modes, all the points have $|c|\ge3.5$ and lie above the angular transition.
For the gravitational modes with $\xi=0.5$, this condition would require $N$ above $7\times10^4$, and we did not compute this sequence.
For the scalar modes, $A(c)-\ii c$ stays within $0.13$ of its limit $-3/4$ for $|c|\ge1.4$, and we impose no such condition.

These calculations use extended precision arithmetic with a 64 bit mantissa, rather than the multiprecision settings of the 24 frequencies.
We compute $A(c)$ by Chebyshev collocation and follow it continuously from $c=0$.
The continued fraction is started at an index $j_0$ that grows with $N$, with the asymptotic tail of Ref.~\cite{Nollert}.
The ratio $j_0/N$ lies between $7.6\times10^2$ and $1.3\times10^5$.
For the scalar modes, we take larger values for smaller $\operatorname{Re}\delta_N$, since the convergence of the continued fraction gets worse there.
The root search starts from the model root $\Omega_N^{(\kappa)}$.
We recomputed all the points with a separate implementation of the same method, using double-double arithmetic where rounding errors matter.
The separate calculation follows each sequence from the Schwarzschild frequency with $n=N$ at $\chi=0$, and it confirms the label $N$.
The two calculations agree within $1.5\times10^{-9}$.
For the scalar modes with $\xi=1.75$ and $2$, the model roots have $\operatorname{Re}\delta_N$ below $3\times10^{-3}$ in this range of $N$, and we did not compute these sequences.
The reason is that $\operatorname{Re}\F_{0,0}$ vanishes at $\xi\simeq1.921$, as shown by Eq.~\eqref{supp:signRe}.

We fit $\delta_N=\Omega_N+\ii w_N$ by polynomials in $\eps=w_N^{-1/4}$.
We use degrees 2, 3, and 4 with all the points, and degrees 2 and 3 with the half of the points at the largest $N$.
The extrapolated value is the median of the five constant terms, taken separately for the real and imaginary parts.
The quoted uncertainty is half of their range.
It is an empirical estimate of the fit dependence rather than an error bound.
The function $\F_{0,s}$ is not used in the fits.

The fits use polynomials in $\eps$, which is an assumption of the extrapolation.
The bound \eqref{supp:rootBound} excludes a term of order $\eps^2\log(1/\eps)$, but it does not establish an expansion in powers of $\eps$.
We tested this choice for the gravitational modes with $\xi=1$, where the sequence extends to $N=6.4\times10^5$.
The ratio $\Delta_{\kappa,N}/\eps^2$ changes from $-0.149+0.107\ii$ at $N=5000$ to $-0.089+0.090\ii$ at $N=6.4\times10^5$.
A quadratic polynomial in $\eps$ reproduces the 24 values with a root mean square deviation of $4\times10^{-5}$ and gives the limit $-0.065+0.086\ii$.
These data agree with the bound and support the polynomial form of the fits.
With $\bar L=0$, the modulus of $(\delta_N-\F_{0,s}-\eps\F_{1,s,m,\bar L})/\eps^2$ stays below $0.3$ along the scalar sequences and between $0.9$ and $1.8$ along the gravitational sequences, in agreement with Eq.~\eqref{supp:frequencyExpansion}.

Table~\ref{supp:tabExtrap} compares the extrapolated values with $\F_{0,s}(\xi)$.
We compute $\F_{0,s}$ in Fig.~\ref{fig:crossover} of the Letter and in Table~\ref{supp:tabExtrap} from the spectral determinant of the sextic oscillator, through Eqs.~\eqref{eq:Q} and \eqref{eq:Tproduct} of the Letter.
Direct integration of the inner equation gives the same values within $3\times10^{-10}$.
The moduli of the differences are below $6\times10^{-4}$ for the scalar modes and below $8\times10^{-4}$ for the gravitational modes.
The differences of the real and imaginary parts are within the quoted uncertainties, with two exceptions.
For the imaginary part of the gravitational modes with $\xi=0.75$, the difference is $0.00054$ and the quoted uncertainty is $0.00027$.
For the imaginary part of the gravitational modes with $\xi=1$, the difference is $0.00064$ and the quoted uncertainty is $0.00060$.
At the largest $N$ of each sequence, $|\Delta_{0,N}|$ is still between $0.005$ and $0.052$.
The ancillary files contain the sequences of this subsection and a script that computes $\F_{0,s}$.

\begin{table}[htbp]
\centering\small
\caption{Extrapolation of $\delta_N=\Omega_N+\ii w_N$ to $N\to\infty$ at fixed $\xi$, compared with $\F_{0,s}(\xi)$. The numbers in parentheses are the uncertainties of the extrapolation in the last digits. They are empirical estimates of the fit dependence.}\label{supp:tabExtrap}
\begin{tabular}{rrrrrrrr}
\hline\hline
$s$ & $\xi$ & Range of $N$ & Points & $\operatorname{Re}\delta_\infty$ & $\operatorname{Im}\delta_\infty$ & $\operatorname{Re}\F_{0,s}$ & $\operatorname{Im}\F_{0,s}$\\
\hline
$0$ & $0.5$ & $2000$--$40000$ & $20$ & $0.0746(9)$ & $-0.0133(10)$ & $0.0746$ & $-0.0134$ \\
$0$ & $0.75$ & $2000$--$40000$ & $20$ & $0.0585(5)$ & $-0.0323(6)$ & $0.0588$ & $-0.0318$ \\
$0$ & $1$ & $2000$--$40000$ & $20$ & $0.0380(4)$ & $-0.0613(5)$ & $0.0383$ & $-0.0610$ \\
$0$ & $1.25$ & $2000$--$40000$ & $20$ & $0.0175(4)$ & $-0.1037(5)$ & $0.0175$ & $-0.1036$ \\
$0$ & $1.5$ & $1000$--$10000$ & $20$ & $0.0043(7)$ & $-0.1577(14)$ & $0.0041$ & $-0.1574$ \\
$-2$ & $0.75$ & $16000$--$480000$ & $23$ & $0.0776(9)$ & $-0.0105(3)$ & $0.0778$ & $-0.0100$ \\
$-2$ & $1$ & $5000$--$640000$ & $24$ & $0.0701(7)$ & $-0.0188(6)$ & $0.0704$ & $-0.0182$ \\
$-2$ & $1.25$ & $5000$--$320000$ & $23$ & $0.0612(7)$ & $-0.0299(9)$ & $0.0610$ & $-0.0294$ \\
$-2$ & $1.5$ & $5000$--$160000$ & $22$ & $0.0501(8)$ & $-0.0446(12)$ & $0.0499$ & $-0.0441$ \\
$-2$ & $1.75$ & $5000$--$40000$ & $20$ & $0.0380(11)$ & $-0.0639(18)$ & $0.0377$ & $-0.0632$ \\
$-2$ & $2$ & $5000$--$40000$ & $20$ & $0.0256(10)$ & $-0.0878(18)$ & $0.0253$ & $-0.0874$ \\
$-2$ & $2.25$ & $5000$--$40000$ & $20$ & $0.0146(9)$ & $-0.1172(16)$ & $0.0144$ & $-0.1170$ \\
$-2$ & $2.5$ & $5000$--$40000$ & $20$ & $0.0064(8)$ & $-0.1513(13)$ & $0.0063$ & $-0.1515$ \\
\hline\hline
\end{tabular}
\end{table}

\subsection{Boundary coefficients away from QNM roots}
We test Eq.~\eqref{supp:matching} by computing $\mathcal D$ and the inner $T_s$ separately, without imposing a QNM condition.
For $\Omega=-\ii w_N+0.10+0.125\ii$, $\chi=\xi w_N^{-3/4}$, and scalar $\xi=1$, the difference $|\mathcal D-e^{4\pi\Omega}-T_0(\kappa_{\rm eff})|$ divided by $\eps^2$ is $6.35$, $6.21$, and $6.05$ at $N=400$, $1600$, and $6400$.
It does not grow, in agreement with Eq.~\eqref{supp:matching}.
A computation through the horizon to infinity connection coefficients and a directly continued outgoing wave give the same value of $\mathcal D$ to about nine digits at the low $N$ control points.

We also test the incoming coefficient identity \eqref{supp:Gfactor} by integrating the standard and dual equations separately.
The gamma multiplier is applied only after obtaining both coefficients.
Five input points cover the convergent Euler domain, its continuation, and scalar, electromagnetic, and gravitational Kerr parameter assignments.
At 65 digits, starting radius 100 and Taylor order 100, all relative differences are below $10^{-49}$.
At 85 digits, radius 140 and order 140, they are below $2\times10^{-69}$.
The angular input in the last three tests is an arbitrary complex value.

\subsection{Local connection and determinant checks}
The standard and dual sextic equations are integrated separately.
We initialize the canonical waves of Eq.~\eqref{supp:Y} with their asymptotic series, integrate inward to unit radius, and continue on a circular path.
The rotated energy, deformation, prefactor and lifted argument are those of Eq.~\eqref{supp:rotated}.
The main and stronger runs use 45 and 65 digits, starting radii 4 and 4.5, asymptotic orders 100 and 160, and Taylor orders 36 and 48.

We use the scaled difference
\begin{equation}
d_{\rm SD}=\frac{|T_s^{\rm T}-T_s^{\rm d}|}{\max(1,|T_s^{\rm T}|,|T_s^{\rm d}|)}.
\label{supp:dSD}
\end{equation}
We test $E=0$, $1$, and $2+\ii$ for $s=-1$, and $E=1$, $2+\ii$, and $\pm4\sqrt3$ for $s=-2$.
In all seven tests, $d_{\rm SD}$ is below $3\times10^{-27}$ in the main run and below $4\times10^{-41}$ in the stronger run.
At the quasi-exactly solvable energies $E=0$ for $s=-1$ and $E=\pm4\sqrt3$ for $s=-2$, both integrations give $T_s=-1$.
These numbers measure the agreement of the two integrations rather than certified digits.

At $E=2+\ii$, the basic multipliers separately reproduce the fused coefficients within $1.14\times10^{-42}$ and $3.52\times10^{-43}$ for $s=-1,-2$.
Omitting the phase gives differences about 1.17 and 2.55.
Symbolic substitution gives zero for the two intertwining residuals and the standard limiting potential, and gives the Wronskian multipliers $E$ and $E^2-48$.

The determinant calculation uses the weighted equation \eqref{supp:weighted} on the positive real axis, without continuing around the origin.
We match regular and decaying endpoint series and normalize the Wronskian by its value at $X=0$.
Evaluations at $X,-X,\ii X,-\ii X$ are inserted in Eq.~\eqref{supp:Tproduct} and compared with direct inner continuation.
The latter uses two rays and a clockwise $3\pi/2$ arc of radius one.
For $s=0$ and $-2$ and $\xi=0$, $0.5$, $1$, $2$, $4$, $6$, and $8$, the two methods agree within a relative difference of $1.1\times10^{-12}$.
Changing the settings of the determinant calculation changes $T_s$ by at most $1.1\times10^{-14}$.
The values at large $\xi$ test the inner equation and do not correspond to physical Kerr roots.

\end{document}